%% file: Grw70.tex
\documentclass[useAMS,fleqn,usenatbib]{mnras}
\usepackage{newtxtext,newtxmath}
\usepackage[T1]{fontenc}
\usepackage{ae,aecompl}
\usepackage{graphicx}	
\usepackage{amsmath}	
\usepackage{amssymb}	
\newcommand{\fo}{\ensuremath{f^\parallel}}
\newcommand{\fe}{\ensuremath{f^\perp}}
\newcommand{\pq}{\ensuremath{P_Q}}
\newcommand{\pu}{\ensuremath{P_U}}
\newcommand{\pv}{\ensuremath{P_V}}
\newcommand{\pX}{\ensuremath{P_X}}
\newcommand{\pl}{\ensuremath{P_{\rm L}}}

\newcommand{\nnx}{\ensuremath{N_X}}

\newcommand{\grw}{Grw+70$^\circ$\,8247}
\newcommand{\bz}{\ensuremath{\langle B_z \rangle}}

\newcommand{\snr}{\ensuremath{S/N}}

\newcommand{\alphatel}{\ensuremath{\alpha_\mathrm{Tel}}}
\newcommand{\deltatel}{\ensuremath{\delta_\mathrm{Tel}}}
\newcommand{\alphasource}{\ensuremath{\alpha_\mathrm{o}}}
\newcommand{\deltasource}{\ensuremath{\delta_\mathrm{o}}}

\title[Polarimetric variability of the white dwarf Grw+70$^\circ$\,8247]{The long term polarimetric
  variability of the strongly magnetic white dwarf Grw+70$^\circ$\,8247}
\author[Bagnulo \& Landstreet]{S.~Bagnulo$^{1}$ and J.D.~Landstreet$^{1,2}$\\
$^{1}$Armagh Observatory and Planetarium, College Hill, Armagh BT61 9DG, UK.
{\rm E-mail: Stefano.Bagnulo@Armagh.ac.uk; John.Landstreet@Armagh.ac.uk}\\
$^{2}$Department of Physics \& Astronomy, University of Western Ontario, London, Ontario, Canada N6A 3K7
{\rm E-mail: jlandstr@uwo.ca}\\
}

\begin{document}

\date{Accepted 2019 April 11. Received 2019 April 10; in original form 2019 February 17}

\pagerange{\pageref{firstpage}--\pageref{lastpage}} \pubyear{2018}

 \maketitle

\label{firstpage}

\begin{abstract}
  Some of the white dwarfs exhibit among the strongest magnetic fields
  in the universe.  Many of these degenerate magnetic stars are also
  rotating very slowly. Among these objects, Grw+70$^\circ$\,8247,
  with its century-long suspected rotation period and its 400\,MG
  magnetic field, stands as a particularly interesting
  object. Surprisingly, for this star, the first white dwarf in which
  a magnetic field was discovered, no spectropolarimetric observations
  have been discussed in the literature in the last 40
  years. Here we present two sets of linear and circular polarisation
  spectra taken in 2015 and 2018, and we compare them with
  spectropolarimetric data obtained in the 1970s.  Polarisation shows
  variability over a time interval of four decades, but some subtle
  changes may have been detected even over a three year time
  interval. Using the variation of the polarisation position angle as
  a proxy for the rotation of the magnetic axis in the plane of the
  sky, we conclude that the star's rotation period probably lies in
  the range of $10^2$ to $10^3$ years. Our data analysis is
  accompanied by a description of our various calibrations and tests
  of the ISIS instrument at the William Herschel Telescope that may be
  of general interest for linear spectropolarimetric measurements. We
  also found discrepancies in the sign of circular polarisation as
  reported in the literature, and made explicit the definitions that
  we have adopted.
\end{abstract}

\begin{keywords}
polarization -- stars: white dwarfs -- stars: magnetic fields -- stars: individuals: Grw+70$^\circ$\,8247
\end{keywords}

\section{Introduction}\label{Sect_Intro}
A small percentage of white dwarfs (WDs) exhibit a magnetic field
organised on a large scale over the stellar surface, probably the
fossil remnant of a field that originated during an earlier stage of
the star's life. The mean magnitude of such fields ranges over more
than five dex of strength, from a few kG to about 1000\,MG, and can
have important physical effects on the host star, for example by
transferring angular momentum internally and exchanging it with
circumstellar material, and through suppressing atmospheric and
envelope convection under many circumstances.

In a larger context, the fields of the magnetic white dwarfs (MWDs) are an
example of the occasional occurrence of large static fields in a
variety of stars during stellar evolution, including upper main sequence
stars, white dwarfs, and neutron stars. The origin of these fields is still
essentially completely mysterious. This phenomenon may be capable of
providing valuable information on specific evolution paths (for
example, of merging binary systems) that at present cannot be
exploited because we have not understood some basic stellar
processes. In addition, the fields of white dwarfs provide access to a
plasma laboratory permeated by fields far too large to be generated
statically in terrestrial laboratories. 

Up to a field strength of the order of 80 or 100\,MG, white dwarf
fields are detected and studied using the splitting, shifting and
polarisation of spectral lines. For higher field strengths, spectral
lines are smeared almost out of recognition but the fields may be
identified through the broad-band continuum circular and linear
polarisation usually observed throughout the optical and ultraviolet
wavelength windows \citep{Feretal15}, as well as through (usually very
shallow) broad spectral features in the flux spectra.

Beyond identifying magnetic fields in white dwarfs, and estimating the
magnitude of the magnetic field, it is of considerable interest to try
to understand the distribution of field strength and direction over
the stellar surface. This kind of information can help to clarify the
interaction of the field with other physical processes in the
underlying star, and may provide clues about the still unclear
processes that lead to the occurrence of magnetic fields in a
small fraction of white dwarfs.

Study of field structure generally requires observing the star
repeatedly as it rotates. A number of magnetic white dwarfs 
exhibit short term variability, with a period that is associated to
the star's rotation period (typically from a few hours to a few
days). As in the case of the better studied magnetic chemically
peculiar stars of the main sequence (Ap stars), the magnetic
variability is explained in terms of a magnetic field not symmetric
about the rotation axis, so that the observer sees a magnetic
configuration that changes as the star rotates. This oblique rotator
model was proposed to explain the polarimetric behaviour of the
magnetic Ap stars by \citet{Stibbs50}, and since then has been the
basic model adopted for any large-scale stellar field that is not
currently sustained by dynamo action.

Only about two dozen of the several hundred known MWDs have been
clearly identified as having very high magnetic fields, above about
100 MG \citep[e.g.][Fig.~5]{Feretal15}. About half of these 100+ MG
stars have emerged from the very large lists of new white dwarfs found
as by-products of the Sloan Digital Sky Survey (SDSS). The other very
high field MWDs were mostly found before the first data release from
the SDSS by a variety of means, including surveys of broad-band
circular polarisation or discovery of remarkable flux spectra (flux
spectra are usually denoted here as Stokes $I$ spectra).

For the SDSS very high field MWDs, in general a single low S/N
(typically in the range of 10 to 30), low resolution ($R \sim 2000$)
spectrum is available, and the characteristic field strength assigned
is usually found by comparing the observed spectrum to spectra from a
large grid of precomputed flux spectra of MWDs of a large range of
$T_{\rm eff}$ and field strength. The field structure assumed for the grid is a
combination of low order multipoles, often dominated by a dipole. The
WD atmosphere is assumed to be pure H, and only the spectrum of H is
included in the model spectra, with spectral line positions taken from
tables of atomic level splitting in very large fields such as those of
\citet{Wunetal85}. The computation of the grid spectra and comparison
with observed $I$ spectra is described in detail by
\citet{Jordan92,Eucetal02} and \citet{Kueletal09}.

This method of modelling provides a first, basic characterisation of
typical field strength and a possible, but not unique, and almost
certainly highly simplified, magnetic field geometry. It takes no
account of the possibility that the field may look rather different as
seen from different sides of the MWD as it rotates, and makes no use
of information that may be contained in the linear and/or circular
polarisation spectra (Stokes components $Q$, $U$, and $V$). The results of
such an analysis provide information about the general magnitude of
the field and possibly of the spread of field strength over the
hemisphere visible at the time that the spectrum was taken, but little
further information.

For most of the dozen or so earlier discovered, and generally
brighter, very high field MWDs, more information is often
available. For example, several Stokes $I$ spectra may have been taken,
or spectra in non-optical wavelength regions such as the UV may be
available, or there may be polarisation spectra (often of circularly
polarised light, Stokes $V$, sometimes of the linear polarisation
components $Q$ and $U$). For four of the 100+ MG field MWDs the rotation
period is known, so data taken during a reasonably short interval
could be assigned to distinct lines of sight to the MWD. When
rotational phase-resolved data are available, models of the surface
magnetic field strength and structure may be considerably better
constrained. An excellent example of the quality of modelling that can
be achieved when several phase-resolved spectra of both Stokes $I$ and $V$
are available is the analysis of PG\,1015$+$014 carried out by \citet{Eucetal06}
on the basis of modelling tools developed by \citet{Jordan92}
and \citet{Eucetal02}. Their results clearly show that the field
of this star is dominated by regions with local field strength mostly
in the range of 70–-80\,MG, and provide fairly well constrained low
resolution maps of the large-scale distribution of flux over the
surface.

However, although models have been developed for each of the four
highest-field MWDs with known rotation periods, none of these four
MWDs have been observed systematically enough to make it possible to
try to model a series of phase-resolved flux and polarisation spectra.

Remarkably, for several of the earliest discovered and most intensely
studied high-field magnetic white dwarfs, including the prototype,
\grw\ = WD\,1900+705\footnote{In addition to being the first magnetic
  white dwarf discovered, and the first high-field MWD, \grw\ (which
  was identified and catalogued in the Greenwich Observatory zone of the
  Astrographic Catalogue) was only the fifth white dwarf ever
  identified \citep{Kuip35}.}, it is not even clear whether any real
variations have been detected. Since any magnetic configuation not
strictly symmetric about the rotation axis of the star should lead to
some kind of observable variability as the underlying star rotates,
these non-varying stars are suspected of having extremely long
rotation periods, of the order of decades or centuries \citep[as is
  also the case for a few middle main sequence magnetic Ap stars,
  e.g.][]{Mathys17}.  For the MWDs in the tranche with fields above
100\,MG, variability may have been detected, but no rotation periods
have been established that are longer than 1.4 days.

For these high-field stars, searching for modest levels of variability
of polarisation through broad-band filters over time spans of decades is not
straightforward. Both filter bandpasses and detector wavelength
sensitivity change with time and technology, and observed variations may
simply be the caused by a different sampling of the stellar 
polarised spectrum.  The
information content of filter polarimetry is also quite limited.

There is little doubt that the most robust method of identifying
variability, measuring the rotation period, and obtaining spectral
information useful for modelling and mapping the field structure, is
to repeatedly obtain polarised spectra through large wavelength
windows (say 2000\,\AA\ or more), using high enough spectral
resolution so that measurements taken over time with different
instruments can be reliably compared. Perhaps surprisingly, this has
so far hardly been done, even though at least two current facility
spectropolarimeters, ISIS at the William Herschel Telescope of the
ING, and FORS at the ESO VLT, are quite capable of measuring both
circular and linear continuum polarisation accurately over large
wavelength ranges, and have been capable of doing this for at least
two decades. Fortunately, for a few of the brightest, most strongly
magnetic and highly polarised white dwarfs, some low resolution
archival spectropolarimetry is available for comparison with modern
data \citep{Angetal72,LanAng74,LanAng75,Angetal85,Schetal96}.

In this paper we present new moderate resolution line and
continuum spectropolarimetry for \grw, and we compare our
data to archival spectropolarimetry.  It has not
yet been established that \grw\ is actually variable, and on the basis
of apparently constant data available decades ago it has been
suggested that this white dwarf may be rotating with a period of order
10$^2$\,yr or more.

The search for subtle variations in the stellar polarised spectra has
prompted us to pay particular attention to the characterisation of the
polarimetric module of the instrument(s) employed in our
observations, so that if differences between new spectropolarimetry
and archival data are found, we can be reasonably sure that they are
not due to instrumental artefacts in the old or the new data.

\section{Definitions of the observed polarisation, with a special
attention to the sign of circular polarisation}\label{Sect_Def}
In this paper we will consider the reduced Stokes parameters $\pX=X/I$
(where $X=Q$, $U$, $V$) and the corresponding null profiles \nnx. It
will also be useful to describe the linear polarisation in terms of
the fraction of linear polarisation $\pl = (\pq^2+\pu^2)^{1/2}$ and
its position angle $\Theta$, such that $\pq=\pl\,\cos(2\Theta)$ and
$\pu=\pl\,\sin(2\Theta)$. The Stokes parameters are defined according
to \citet{Shurcliff62}, and for linear polarisation we haved adopted as a
reference direction the great circle passing through the object and the
North Celestial Pole. The position angle of the linear polarisation
$\Theta$ is measured positive looking at the source, counting
counterclockwise from the great circle passing through the object and
the North Celestial pole.

The two representations of linear polarisation ($\pq,\pu$) and
($\pl,\Theta$) are equivalent, but from the physical and geometrical
point of view, $(\pl,\Theta)$ may be more convenient for a physical
understanding of variability and discrepancies. In case of the
observations of standard stars, an incorrect alignement of the
polarimetric optics is promptly detected by a discrepancy in the
position angle of the polarisation. The position angle of polarisation
due to the presence of a magnetic field is directly influenced by the
transverse components of the magnetic field; for instance, in the
Zeeman regime, and neglecting the magneto-optical effects, the
direction of the maximum polarisation is parallel to the projected
magnetic axis \citep[e.g.][]{Lanetal93}.  For small polarisation
values, the position angle is not well defined, and may be
dramatically affected even by small amounts of instrumental
polarisation or other spurious effects; therefore sometimes the
representation ($\pq,\pu$) will be more convenient. In this paper, to
report our observations of unpolarised standard stars, we will adopt
the ($\pq,\pu$) representation; to report our observations of standard
stars for linear polarisation we will adopt the ($\pl,\Theta$)
representation.  For our primary target, which is a magnetic star but
has a polarisation close to zero at certain wavelength ranges, we will
describe the linear polarisation both with $(\pq,\pu)$ and with
$(\pl,\Theta)$.

Our main target, \grw, is usually thought of as a star with circular
polarisation approximately constant with time (and as such it was
often suggested as a standard star for circular
polarisation). However, it is found that literature data of circular
polarisation measured in similar interval wavelength ranges have
sometimes discrepant signs. For instance, \citet{Kemetal70} and
\citet{AngLan70b} reported a positive sign; \citet{LanAng75} and
\citet{Butetal09} a negative value. This sign ambiguity is probably
related to the adoption of different definitions of the Stokes
parameters, and in particular of the sign of circular polarisation, as
discussed for instance by \citet{Clarke74}, \citet{Lanetal07} and
\citet{Bagetal09}.

In this work we adopt the definition given in \citet{LanLan04},
\citet{Lanetal07} and \citet{Bagetal09}, and consistent with \citet{Shurcliff62}, i.e.,
circular polarisation is {\it positive} (or {\it right-handed}) when
the tip of the electric field vector is seen to rotate {\it clockwise}
in a fixed plane perpendicular to the propagation of the light,
looking at the source. As explained in \citet{BagLan18}, we determine
the sign of our polarisation measurements by measuring the
longitudinal field of well known magnetic stars using the weak-field
relationship \citep[e.g.][]{Bagetal02}
\begin{equation}
  \frac{V}{I} = -g_{\rm eff} \ 4.67\,10^{-13}\
  \lambda^2\ \frac{1}{I} \frac{{\rm d}I}{{\rm d}\lambda} \bz
  \label{Eq_Bz}
\end{equation}
which tells us that a positive longitudinal field will cause Stokes $V$ to be
positive in the blue wing of an absorption line and negative in the
red wing. Some magnetic Ap stars that have been found with constant
polarity over decades may be used as reference stars; for instance
HD\,94660 and $\gamma$\,Equ (since the 1970s) are always reported with
a negative longitudinal field, and our field measurements of
$\gamma$\,Equ are indeed consistent with a negative sign (see
Sect.~\ref{Sect_CirAli}).  Of course our empirical approach to link the sign
of circular polarisation to the sign of the magnetic field just shifts
the problem into the original determination of the polarity of the
magnetic field of well known magnetic stars, which probably goes back
to the extensive work on Ap stars by \citet{Babcock58}.

The ambiguity in the sign of circular polarisation does not seem to
have affected the sign of the magnetic field reported in the
literature.  For instance, we note that \citet{Putney95} reports a
negative circular polarisation for \grw\ (see her Fig.~1c), i.e., with
the sign opposite to what we measure. However, for the field determinations of MWDs in
the Zeeman regime, she adopts Eq.~(\ref{Eq_Bz}) with the plus sign in
its right hand term, and so other authors do \citep[e.g.][]{BorLan80}.
This suggests that the sign of stellar magnetic fields are
probably defined consistently in most if not all literature.

Removing the ambiguity in the sign of circular polarisation is
obviously crucial to allow one a proper comparison of the observations
of different authors and epochs. Establishing the geometrical
significance of the sign of the circular polarisation is also
important when the original of circular polarisation is associated to
homochirality.

\section{New Observations: instrument settings}
We carried out our new spectropolarimetric observations with the ISIS
instrument of the William Herschel Telescope. The ISIS polarimetric
module consists of an achromatic retarder waveplate ($\lambda/2$ for
observations of linear polarisation, or $\lambda/4$ for observations
of circular polarisation) which can be rotated to a series of fixed
positions, followed by a Savart plate that splits the incoming
radiation into two beams linearly polarised in directions perpendicular to each
other, one along the principal plane of the plate (the parallel beam
\fo), and one perpendicular to that plane (the perpendicular beam
\fe). These beams propagate parallel to each other but separated. A
special 18\arcsec\ dekker prevents the superposition of each beam
split by the Savart plate with the light coming from the other parts
of the observed field of view. The use of a dichroic beam splitter
with cut-off centred at 5300\,\AA\ allows us to observe
simultaneously in the blue arm with grating R600B (which we have set
at a central wavelength = 4400\,\AA\ so as to cover the spectral
range of 3700--5300\,\AA), and in the red arm with grating R1200R
(central wavelength = 6500\,\AA, to cover the spectral range
6100--6900\,\AA). Our observations were obtained with slit width
of 1.0\arcsec, 1.2\arcsec\ and 2.0\arcsec\ for a spectral resolving power $R$ 
of 2500, 2050 and 1280 in the blue arm, and 8950, 7100 and 4500 in the
red arm, respectively. Spectral features of \grw\ are broad enough that
observations obtained at the lowest resolution with the 2\arcsec\
slit width did not appear more smeared than those obtained
at the highest resolution with a 1\arcsec\ slit width.

For \grw\ we have obtained three observations of circular
polarisation, two in August 2015 and one in November 2018, and two
observations of linear polarisation, one in August 2015 and one in
November 2018.  During the November 2018 run (night 21 to 22),
we obtained an additional set of observations both in circular and linear polarisation,
but the selection of a wrong combination of decker and window readout
did not allow us to record the background. Therefore, both datasets
obtained on the night of November 21 were discarded.

For our measurements we adopted the use of the beam swapping technique
\citep[e.g.][]{Bagetal09} which minimises the impact of instrumental
polarisation. ISIS circular polarisation
measurements were obtained using a series of four exposures with the
$\lambda/4$ retarder waveplate set at position angles $-45^\circ$,
$+45^\circ$, $+45^\circ$, $-45^\circ$, while our
linear polarisation measurements were obtained with a series of eight
exposures with the $\lambda/2$ waveplate set 
at position angles $0^\circ$, $22.5^\circ$, $45^\circ$,
$67.5^\circ$, $90^\circ$, $112.5^\circ$, $135^\circ$ and
$157.5^\circ$.

The redundancy of our observing strategy allows us to measure also
the so called null profiles \nnx, which give us an experimental
estimate of the uncertainties \citep{Donetal97,Bagetal09}.

The dates and times of the new observations, along with other data to be explained later, are listed in Table~\ref{Table_Grw}. 

\section{Data reduction}\label{Sect_DR}

Reduction of circular polarisation data is fully explained in
\citet{BagLan18} and references therein. Most of the steps used to
reduce linear spectropolarimetric data are similar to the procedure
for reducing circular polarisation data, but additional steps have to
be taken to correct linear polarisation measurements for the
chromatism of the retarder waveplate. Also, because sky background may be
linearly polarised, its subtraction is particularly crucial.
These steps will be described in Sect.~\ref{Sect_Chroma} and
\ref{Sect_BKG} below.

\subsection{Correction for the chromatism of the retarder waveplate}\label{Sect_Chroma}
The position angle of the optical axes of the retarder wavelate is
wavelength dependent, therefore the measured position angle will be
affected by a (wavelength-dependent) offset $\epsilon (\lambda)$,
which may be estimated by measuring the polarisation angle of a source
for which the polarisation position angle is well known.  Following
\citet{Bagetal09}, the reduced Stokes parameters were obtained from
the observed $\pq'$ and $\pu'$ using
\begin{equation}
  \begin{array}{rcl}
  \pq(\lambda) &=& \phantom{-}\pq'(\lambda)\,\cos(2\epsilon(\lambda)) + \pu'\sin(2\epsilon(\lambda))\\
  \pu(\lambda) &=&          - \pq'(\lambda)\,\sin(2\epsilon(\lambda)) + \pu'\cos(2\epsilon(\lambda))\\
  \end{array}
  \end{equation}
To calculate $\epsilon(\lambda)$ we have used observations of asteroid
(2) Pallas that was observed by one of us (SB) on 2014-05-12 with ISIS in
spectropolarimetric mode.

\subsection{Background subtraction}\label{Sect_BKG}
Our observations of \grw\ were obtained close to Full Moon. For a
$V=13.5$ target, the background level is relatively small, but because it
is highly polarised, it may introduce spurious effects if not
correctly subtracted (espcially in the case of linear polarisation measurements).
In both our linear polarisation measurements of
\grw, the background level was about 10\,\% of the source level, both in
the blue and in the red spectral range. During the 2015 observations,
the background polarisation was $\sim 36$\,\% in the blue and $\sim
22$\,\% in the red. In the 2018
observations, the background polarisation was about 51\,\% in the blue and
28\,\% in the red (see Sect.~\ref{Sect_BKG_Sky}).

In ISIS, the background sky may be measured in four 5\arcsec\ strips (two
per beam) parallel to the strips illuminated by the target, each
separated by 18\arcsec\ from the central beam which is centred on the
target \citep[see the bottom panels of Fig.~1 of][]{BagLan18}.

We have experimented with different algorithms for background
subtraction among those that are offered by the IRAF procedure {\tt
  apall}, finding that our final results for the polarisation of
\grw\ were always consistent with each other within photon noise error bars.
We have also carried out a more critical experiment as follows:
instead of extracting the background from both full side strips, we
used only a small window of one of the strips (of course still
considering both beams split by the Savart plate). The results of this
experiment were fully consistent (within error bars) with those
obtained by estimating the background flux using both full strips.
We are therefore satisfied that the presence of a non-negligible
polarised backgrounds does not affect our data, e.g., by introducing
a systematic effect, apart from reducing their \snr.

In contrast, our linear polarisation measurements obtained
during the rejected night 2018-09-21 could not be background subtracted, and
the measured values were totally inconsistent with those obtained on the
next night, when background subtraction could be correctly performed.

For circular polarisation, sky subtraction is a less critical step
because the sky is not circular polarised. The effect of not subtracting
the backround would be a dilution of the polarisation signal by a factor
roughly equal to the ratio between the sky flux and the star flux.

\subsection{From spectropolarimetry to broadband polarimetry}\label{Sect_BBP}
Most of literature polarimetric data were obtained in broadband filters. For
comparison purposes, it may be useful to compute synthetic broadband
values using the formula
\begin{equation}
  \pX(F) = \frac{\int_{0}^{\infty}\mathrm{d}{\lambda}\ \pX(\lambda)\,
    I_X(\lambda)\, T_{\rm F}(\lambda)}{\int_{0}^{\infty}\mathrm{d}\lambda\ I_X(\lambda)\,T_{\rm F}(\lambda)}
\label{Eq_BBP}
\end{equation}
where $T_{F}$ is the transmission function of the $F$ filter and $I_x$
is the sum of the fluxes in all beams and all images used to obtain
the reduced Stokes $X$ parameter.  The setting in the blue ISIS arm
that we have used for the observations of the magnetic white dwarfs
covers the $B$ filter, and the setting in the red ISIS covers the
spectral range of the $R$ filter. After integrating the signal over a
large wavelength interval, photon noise error becomes totally
neglibile, but systematics and other effects are still present. As
tentative estimates of the uncertaties of the broadband polarisation (BBP)
values we will adopt the values obtained by replacing \pX\ in
Eq.~(\ref{Eq_BBP}) with the corresponding null profiles \nnx, with the
caveat that this method still leads to an underestimate of the errors,
as it would not include, for instance, telescope and instrumental
polarisation, or errors in matching the original filter transmission
function or detector wavelength sensitivity function.

\section{Quality checks}\label{Sect_QC}

In the following we investigate the instrument accuracy and
stability. We examine various potential issues that may affect
the accuracy of our measurements: cross-talk from $I$ to $V$
(Sect.~\ref{Sect_IPcir}), from $I$ to $Q$ and $U$
(Sect.~\ref{Sect_IPlin}), and from $Q$ and $U$ to $V$
(Sect.~\ref{Sect_Xtalk}). We present observations of magnetic stars and standard stars
for linear polarisation that we have performed to check the correct
alignment of the polarimetric optics (Sect.\ref{Sect_CirAli} for
circular polarisation measurements and Sect.~\ref{Sect_LinAli} for
linear polarisation measurements).  In Sect.~\ref{Sect_DIC} we
discuss whether the use of a dichroic introduces a signal of spurious
polarisation, as warned by the ISIS user manual. In Sect.~\ref{Sect_Short}
we discuss the polarimetric efficiency at shorter wavelengths. All results
are then summarised in Sect.~\ref{Sect_Summary}.


\subsection{Estimate of instrumental circular polarisation in the continuum}\label{Sect_IPcir}
The large majority of nearby WDs have a spectrum that is intrinsically
unpolarised. Our extensive surveys of WDs \citep[see][]{BagLan18} offer abundant
material to determine whether the instruments that we have used introduce a
spurious signal of circular polarisation. For ISIS, we have found that the signal
of circular polarisation in the continuum is generally $\la 0.03$\,\%
in the blue arm and $\la 0.05$\,\% in the red arm.

\subsection{Observations of unpolarised stars in linear polarisation}\label{Sect_IPlin}
\input{Table_ZeroPol}
\begin{figure}
  \includegraphics*[width=9cm,trim={0.5cm 1.2cm 1.0cm 1.7cm},clip]{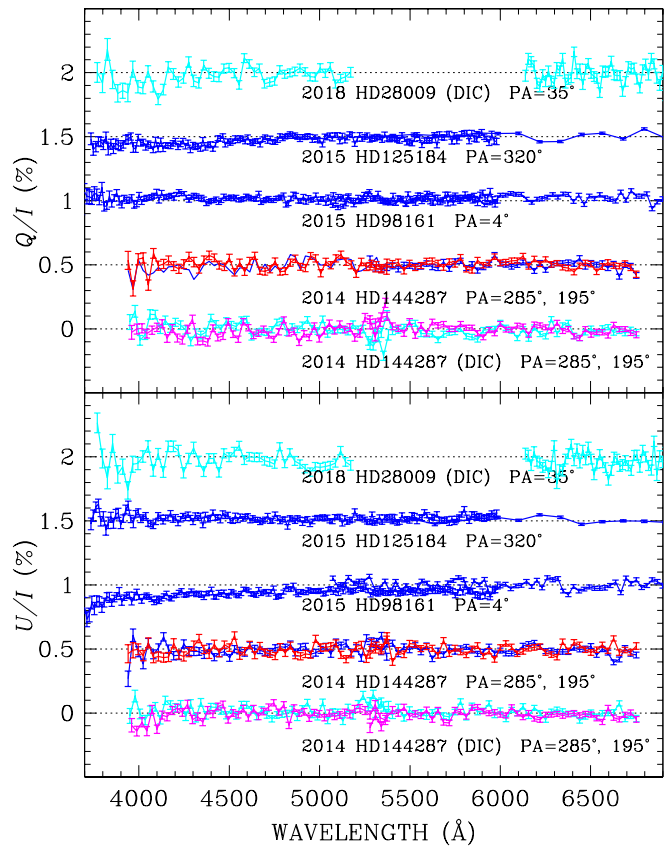}
  \caption{\label{Fig_Zero} Polarisation spectra of unpolarised stars, offset
    in 0.5\,\% steps for display purpose.
    Observations labeled with ``DIC'' (represented with light blue and
    magenta colour) were obtained in both arms simultaneously using a dichroic beam splitter.
    The remaing observations (red and blue solid lines) were obtained
    independently in the two instrument arms. 
    The spectra of HD\,144287 were obtained with the instrument position angle
    at $195^\circ$ (magenta and red lines) and $285^\circ$ (blue and light blue lines).}
  \end{figure}
During the course of various observing runs with ISIS in
spectropolarimetric mode we have obtained a number of measurements in
linear spectropolarimetric mode of unpolarised stars listed in Table~\ref{Table_ZeroPol}
(in our list we have included star HD\,98099 that being within 50\,pc was deemed as
probably non significantly polarised, although was never adopted as unpolarised
standard star).
Some observations have been obtained independently both in the red and in
the blue arm (one after the other), and some have been obtained
simultaneously in both arms with the use of a dichroic beam splitter.
We note that using the highly-sensitive PlanetPol instrument,
\citet{Houetal06} estimated that the spurious polarisation introduced
by the telescope optical surfaces of the WHT is of the order of $1.5\,10^{-5}$.  Assuming that this
was still true during our observing run, any measured signal of
spurious polarisation would come from the ISIS instrument.

\citet{Houetal06} argued that spurious linear polarisation introduced
by the instrument could be characterised with the help of observations
obtained with the instrument set at different position angles on sky. 
For instance, denoting with $P_Q^{(\alpha)}$, $P_U^{(\alpha)}$ the
reduced Stokes parameters measured with the instrument position at PA=
$\alpha$ on sky, one should measure:
\begin{equation}
\label{Eq_Rot}
\begin{array}{rcl}
  \pq^{(0)}  &=&\phantom{-}q + \pq^{\rm instr}\\
  \pq^{(90)} &=&          -q + \pq^{\rm instr} \\
  \pu^{(0)}  &=&\phantom{-}u + \pu^{\rm instr} \\
  \pu^{(90)} &=&          -u + \pu^{\rm instr} \\
\end{array}
\end{equation}
where $q$ and $u$ are the reduced Stokes parameters intrinsic to the source. In general,
for an arbitrary PA $ =\alpha_0$,
\begin{equation}
\begin{array}{rcl} 
\pq^{\rm instr} &=& \frac{1}{2}\, \left(\pq^{(\alpha_0)} + \pq^{(\alpha_0+90)}\right)\\
\pu^{\rm instr} &=& \frac{1}{2}\, \left(\pu^{(\alpha_0)} + \pu^{(\alpha_0+90)}\right) \, . \\
\label{Eq_Instr}
\end{array}
\end{equation}

Our measurements are summarised in Table~\ref{Table_ZeroPol} and in
Fig.~\ref{Fig_Zero}, and show clearly that instrumental polarisation
seems consistently well within 0.1\,\% in both Stokes parameters $Q$
and $U$, with the exception of one observations of HD\,98161.
Star HD\,144287 was observed twice with the instrument position
angle rotated by 90\degr.  Instrumental polarisation measured
via Eq.~(\ref{Eq_Instr}) was found $\la 0.05$\,\% at all wavelengths.

Star HD\,144287 was observed first with the dichroic, then
independently without dichroic in the two arms. A comparison between
the results shown no evidence that the use of dichroic leads to a
spurious signal of linear polarisation in stars that are intrinsically
unpolarised. We will come back to this point in Sect.~\ref{Sect_DIC}.

\subsection{Crosstalk from linear to circular polarisation}\label{Sect_Xtalk}
The optics that precede the retarder waveplate may transform linear
into circular polarisation or viceversa.  Observations of circular
polarisation of a source that is strongly linearly polarised may be
affected by cross-talk from linear to circular polarisation and may
give a non null measurement even if the source is not intrinsically
circularly polarised. This is a well known instrumental effect present
for instance in the FORS instrument of the ESO VLT, in which it is believed
that the instrument collimator (located {\em above} the polarimetric optics) transforms a few per cent of the linear
polarisation into circular polarisation \citep[see Sect.~7.4
  of][]{Bagetal09}. To check whether ISIS at the WHT (which has no dichroic optics above the wave plates) suffers from this
problem, we have observed HD\,25443, a standard star for linear
polarisation, in linear and in circular polarisation, both in the red
and in the blue arm simultanesouly, with the dichroic.  We measured
$\sim 5$\,\% of linear polarisation (consistently with previous
literature), 0.02\,\% of circular polarisation in the blue and less
than 0.05\,\% in the red. Hence cross-talk from linear to circular
polarisation is smaller than 1\,\%. Because the typical linear
polarisation of strongly magnetic white dwarfs is only a few percent,
this crosstalk is an unimportant contributor to the error budget of
$V/I$ measurements.

\subsection{Alignment of the optics for circular polarisation:
  observations of magnetic stars}\label{Sect_CirAli}

The correct alignment of the polarimetric optics used to measure
circular polarisation was checked by measuring stars with known
magnetic fields. For the ISIS observations obtained in our 2015 run,
this check is described by \citet{BagLan18}. In our September 2018 run
(on 2018-09-21 at 20:11 UT) we observed the magnetic Ap star
$\gamma$\,Equ (=\,HD\,201610) and measured a mean longitudinal field
of $-970\pm30$\,G in the blue arm and $-974\pm26$\,G in the red
arm. These values are perfectly internally consistent, and are
consistent with what is expected for that long period ($P \ga 100$\,y)
magnetic variable \citep[see Sect.~5.1.3 and Table~A.1 of][]{BagLan18}.
\input{Table_STD}

\subsection{Alignment of the optics for linear polarisation}\label{Sect_LinAli}
Since we aim to detect subtle variations of the polarimetric
properties of the target star, we need to pay special attention the correct
alignment of the polarimetric optics. This was checked with the help
of measurements of standard stars for linear polarisation (Sect.~\ref{Sect_SSLP}),
of the Moon  (Sect.~\ref{Sect_Moon}), of the twilight sky (Sect.~\ref{Sect_TWL})
and of the background sky during our science observations (Sect.~\ref{Sect_BKG_Sky}).

\subsubsection{Observations of standard stars for linear polarisation}\label{Sect_SSLP}
The most obvious check to do is to observe standard stars for linear
polarisation and compare the results with literature data. In order to
assess the overal instrument stability we have considered ISIS data
obtained in a period of time more extended than the epochs of the
science observations presented in this paper. Since literature data
report measurements in broadband filters, we have used Eq.~(\ref{Eq_BBP})
to integrate the spectra obtained in the blue arm with the $B$ filter response curve,
and the spectra in the red arm with the $R$ filter response curve.
In Table~\ref{Table_STD} we report all our observations of standard stars,
even those that we suspect are affected by spurious signals.
Our objective is not to establish a list of standard stars but to
characterise the instrument and in particular check its stability.
Results that are slightly off from expectations
are still useful ot help to evaluate what can go wrong during
spectopolarimetric observations.

It appears that, apart from the cases flagged in the Table footnotes,
the position angle of the polarisation of standard stars is always
within 1\degr\ of the literature data, and polarisation values
differed at most by 0.1\,\%. In the context of this investigation,
these discrepancies are not significant, our conclusion is that ISIS
observations of standard stars for linear polarisation confirm that
the instrumental systematics are at most 0.1-0.2\,\% and $1 -
2^\circ$. When not dominated by photon noise, uncertainties should be
generally of that order of magnitude.

\subsubsection{The position angle of the polarisation of the
 Moon}\label{Sect_Moon}
The light scattered by the atmosphere-less objects of the solar system
is polarised in the direction either parallel or perpendicular to the
scattering plane. The angle $\Phi$ between the scattering plane and
the great circle passing through the region of the sky pointed by the
telescope and the North Celestial Pole can be obtained from the
relationship \citep{Bagetal06}
\begin{equation}
 \label{Eq_Dir}
 \begin{array}{l}
 \sin \deltatel \cos(\alphasource - \alphatel) = \\
 \ \ \ \cos(\deltatel) \tan(\deltasource) - \sin(\alphasource - \alphatel) \frac{1}{\tan(\Phi)}\; ,\\
 \end{array}
\end{equation}
where $(\alphasource, \deltasource)$ represent the coordinates of the Sun
and $(\alphatel, \deltatel)$ the coordinates of the observed target,
in this case the Moon.

The Moon was observed at the beginning of each night of our
observations. On 2015-09-01 and 2018-11-22, the Moon phase-angle (the
angle between the Sun, the Moon and the observer, not to be confused
with the angle $\Phi$ which identifies the position of the scattering
plane in the plane of the sky) was 25\degr\ and 20\degr. At these
phase-angles, the (small) lunar polarisation is expected to be
perpendicular to the scattering plane, and our measurements were found
within  1-2\degr\ of the expected value $\Phi+90\degr$.

\subsubsection{Position angle of the polarisation of the twilight sky}\label{Sect_TWL}
Assuming a single scattering mechanism, the light scattered by the sky
should be polarised in the direction perpendicular to the
scattering plane.  We verified that the position angle of the
polarisation of the twilight sky was consistent with the value of
$\Phi+90\degr$, where $\Phi$ is again obtained from Eq.~(\ref{Eq_Dir}), in
which $(\alphasource, \deltasource)$ are the coordinates of the Sun and
$(\alphatel, \deltatel)$ are the coordinates of the point on the sky
pointed by the telescope.

\subsubsection{Position angle of the polarisation of the background sky}\label{Sect_BKG_Sky}
The polarisation of the background sky of science observations may be
accurately measured if exposures are sufficiently long and/or
background intensity is sufficiently high, for instance because of the
presence of the Moon.  One can compare the direction of the
polarisation of the background sky and verify that it is perpendicular to the scattering
plane. During our science exposures, the background was illumniated (and highly polarised)
because of the presence of a nearly full Moon (see Sect.~\ref{Sect_BKG}). We
found that the direction of the background sky was oriented within a
few degree of the value $\Phi+90\degr$ as obtained from Eq.~(\ref{Eq_Dir}), where
$(\alphasource, \deltasource)$ are the coordinates of the Moon and $(\alphatel, \deltatel)$
the coordinates of \grw.

\subsection{Does the dichroic produce a spurious polarisation signal?}\label{Sect_DIC}

\begin{figure}
  \includegraphics*[width=9cm,trim={0.5cm 5.7cm 1.0cm 3.0cm},clip]{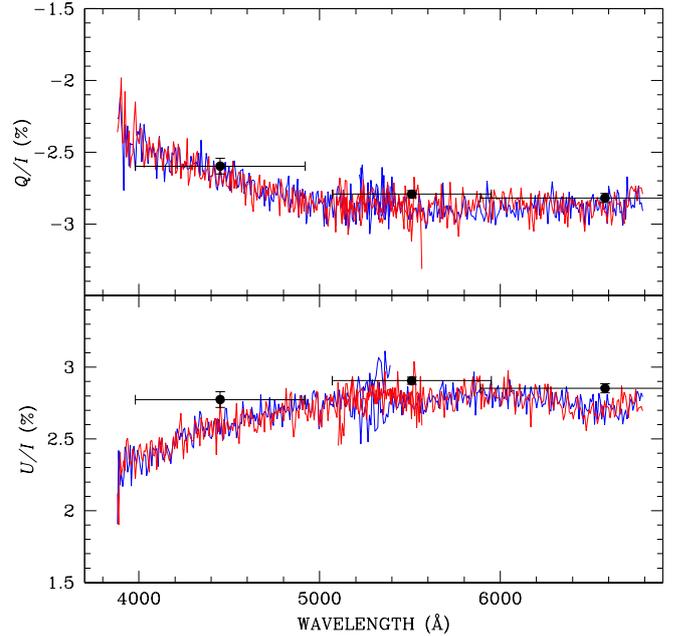}
  \caption{\label{Fig_Dic} Polarisation spectra of star HD\,161056 obtained
    on 6 march 2014 using the dichroic (blue lines) and observing in the
    two arms separately, without the dichoric (red lines).
    Solid circles represent the broadband linear polarisation values
    from \citet{Schetal92} (horizontal
    errorbars corresponds to the filter FWHM). }
  \end{figure}
The ISIS spectrographs has two different camera arms, one optimised for the
red and one for the blue. The insertion of a dichroic beam splitter
allows one to observe simultaneously with both arms. This is a very
valuable instrument feature that doubles the diagnostic capabilities
compared to a situation in which the two arms are fed individually in
separate exposures. The instrument web pages suggests that light
scattered by the back of the dichroic may compromise polarimetric
measurements, and that therefore observations in spectropolarimetric
mode should not be carried out in both arms simultaneously. In
Sect.~4.2.2 of \citet{BagLan18} we found that magnetic field
measurements obtained with the dichroic inserted in the optical path
do not differ significantly from those obtained when the blue and red
arm are fed separately. Table~\ref{Table_STD} shows that our
observations of linear polarisation obtained with the dichroic are
consistent with literature data, suggesting that spurious effects
introduced by the dichroic, if present at all, are probably not very
significant. Moreover, in 2015, the standard star for linear
polarisation HD\,160529 was observed with the dichroic in the blue and
red arms, and also (quasi-simultaneously) in the blue and red arm
individually (see Fig.~\ref{Fig_Dic} and Table~\ref{Table_STD}). A
similar observation was carried out on 6 March 2014 on the standard
star HD\,160556 and on the unpolarised star HD\,144287 (see
Table~\ref{Table_ZeroPol}). The results of these two experiments do not
point to spurious effects introduced by the dichroic. Therefore, even
if the possibility that scattered light may affects polarimetric
measurements should not be forgotten, we proceed assuming that the
measurements obtained with the dichroic are not affected by
significant spurious effects.

\subsection{Polarimetric behaviour at shorter wavelenghts}\label{Sect_Short}
\begin{figure}
  \includegraphics*[width=8.3cm,trim={0.95cm 5.1cm 1.7cm 2.8cm},clip]{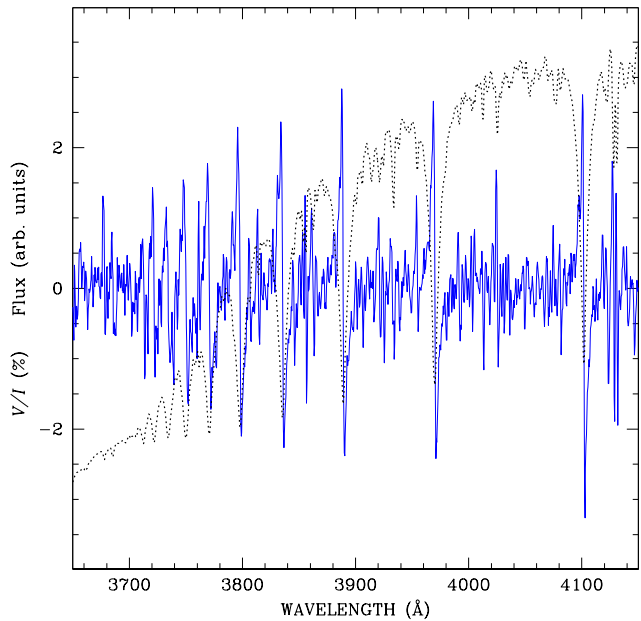}\ 
  \caption{\label{Fig_Ap} The polarised spectrum of the magnetic Ap star HD\,215441 shows
  that the optical properties of ISIS are stable at shorter wavelenghts.}
  \end{figure}
In Sect.~\ref{Sect_Change} we will see that at $\lambda \la
4000$\,\AA, the ISIS circular polarisation spectrum of \grw\ shows
remarkable differences compared to previous data obtained in the 1970s.
To explore whether the observed changes are due to an instrumental
effect (for instance a dramatic change of the retardation or the
position angle of the fast axis of ISIS retarder waveplate), we have
inspected the polarisation spectrum of the strongly magnetic Ap star
HD\,215441 that we obtained in 2015 \citep{BagLan18} with the same
setting as \grw\ (see
Fig.~\ref{Fig_Ap}).  Visual inspection shows that the higher order
Balmer lines appear polarised in a similar way as H$\gamma$ and
H$\beta$ (taking into account the fact that Zeeman effect varies as
$\lambda^2$).  In a more quantitative way, the longitudinal magnetic
field calculated from each individual Balmer line via
Eq.~(\ref{Eq_Bz}) gives the same result (within error bars). We
conclude that, in the observed spectral range, the chromatism of the
quarter retarder waveplate is not reponsible for obvious artefacts.

\subsection{Summary}\label{Sect_Summary}
To summarise the results of this Section: 

\begin{itemize}
\item We find negligible zero point polarisation present in
  observations of both circular and linear polarisation made with ISIS
  in spectropolarimeteric mode.
\item We find negligible crosstalk between Stokes components $Q$, $U$, and $V$ in ISIS data.
\item Both the scale and position angle deduced from ISIS linear spectropolarimetry are extremely accurate.
\item Our methods for correcting for polarised night sky contaminating spectra of faint objects are robust and reliable. 
\item We did not find any evidence that the use of a dichroic beam-splitter with cut-off at 5300\,\AA\ produces a
  signal of spurious polarisation.
\item We did not find evidence for artefacts due to the chromatism of the quarter waveplate.
\end{itemize}

\section{Observations of Grw+70$^\circ$\,8247}\label{Sect_GR}

In the late 1960s James Kemp, an experimental physicist at the
University of Oregon, realised that thermal emission from an
incandescent body in a very strong magnetic field should display
broad-band (spectrally diffuse) circular polarisation, at a level of
the order of $10^{-2}$\,\% for a field of $10^5$\,G
\citep{Kemp70b}. He showed experimentally that this idea is
qualitatively correct \citep{Kempetal70}. When he heard from George
Preston of the Hale Observatories that efforts were being made to
detect large fields in white dwarf stars \citep{Pres70,AngLan70a}, he
adapted his laboratory polarimeter for astronomical observations and
started a search for circular polarisation due to magnetic fields in bright WDs on the 24-inch
telescope of the University of Oregon's Pine Mountain Observatory.

At the suggestion of one of us (JDL), Kemp observed \grw\, and
promptly detected strong circular polarisation.  Following a telephone call from Kemp,
three confirming measurements were obtained by Angel and Landstreet at
Kitt Peak National Observatory, using the photoelectric polarimeter described
by \citet{AngLan70a}. These measurements showed circular polarisation
increasing with wavelength from about 1.5\,\% at 380\,nm to 3.3\,\% at
620\,nm. These observations represented the first discovery of a
magnetic field in a white dwarf \citep{Kemetal70}.

This result triggered a number of further surveys for magnetic fields
in WDs, which up to the present have led to the identification of
several hundred MWDs \citep{Feretal15}.  \grw\ still has one of the
strongest WD magnetic fields known. It is often considered to be a
star with a constant signal of circular polarisation. However, it has
not been monitored nearly as frequently as one could have imagined. In
the following we will review the polarimetric measurements that are
found in the literature (Sect.~\ref{Sect_GrCir} and \ref{Sect_GrLin}),
before presenting our new data (Sect.~\ref{Sect_Our_Observations}) and
comparing them with earlier datasets (Sect.~\ref{Sect_Change}).

\subsection{Observations of circular polarisation in the literature}\label{Sect_GrCir}
In this and in the following Sections we will report the sign of
circular polarisation using our definition of
Sect.~\ref{Sect_Def}. Therefore, some numerical values may appear with
the opposite sign with respect to that found in the original papers.

After the discovery by \citet{Kemetal70}, \citet{AngLan70b} reported a
number of measurements of circular polarisation of \grw\ obtained between 20
June and July 7 1970, made using the photolectric polarimeter descibed
by \citet{AngLan70a}. No on-sky calibration sources were known
for circular polarimetry, and the instrument performance was
tested using circular polarisers.  Observations were obtained without
any filter, covering the spectral range 400 to 700\,nm \cite[see
  Table~1 and Fig.~1 of][]{AngLan70b}.  The circular polarisation was
found to be constant with time over an interval of days.

Circular polarisation was also measured in eight different wavelength
windows from 310 to 800\,nm, using filters with estimated FWHM between
30 and 120\,nm \citep[see Table~2 and Fig.~2 of][]{AngLan70b}.  These
data provided a very low resolution ($R \sim 10$) spectrum of the
wavelength dependence of the circular polarisation of \grw.  The peak
circular polarisation was about 3.7\,\%, around 415\,nm.

\citet{KemSwe70} measured circular polarisation in the infrared, but
we will not follow up further on this spectral region.

\citet{Angetal72} reported further polarimetric monitoring of the
star, which was observed again during four observing runs in late 1970
and early 1971 in circular polarisation, using a variety of
filters. These measurements are compared with the earlier filter
polarisation data.  They appear to rule out
substantial changes on a time scale of months shortward of 600\,nm.
\citet{Angetal72} noted a possibly real variation of circular
polarisation at $\lambda = 740$\,nm, where circular polarisation
decreased from about $2.4$\,\% in 1970 to about $1.0$\,\% in 1972, but
this could have been due to the different red sensitivity of two
different photomultipliers, one with S-20 red response and one with
GaAs photocathodes, used in the polarimeter during this period.

The spectrum of circular polarisation was measured in June 1971 with
higher spectral resolution (resolution 80\,\AA\ below about 5500\,\AA,
and 160\,\AA\ above, or resolving power $R \sim 40 - 50$), using the
multichannel spectrophotometer (MCSP) on the 200-inch Hale telescope
on Mount Palomar, which was converted into a spectropolarimeter with
the addition of a Pockels cell waveplate followed by a polariser in
front of the entrance aperture \citep{Angetal72}. Globally this
polarisation spectrum is in agreement with the very low-resolution
spectrum found with filters, but it has high enough $R$ to begin to
reveal spectral features in Stokes $V/I$ associated with weak
absorption features in the Stokes $I$.

\citet{LanAng75} reported further observations using the MCSP as a
spectropolarimeter.  The circular polarisation spectrum observed in
August 1972 was compared to a new, very similar $V/I$ spectrum from
June 1973. Both were taken with spectral resolution of 80\,\AA\ in the
blue and 160\,\AA\ in the red. No strongly significant differences
among the three MCSP $V/I$ spectra were detected, suggesting that any
changes occur on a time of a decade or more.

One more circular polarisation spectrum was taken in 1976 in the
spectral window 4--7000\,\AA. This spectrum, which has strongly
variable resolution decreasing towards the red, was obtained using a
prism spectropolarimeter with a Digicon detector on the Steward
Observatory 2.3-m telescope \citep{Angetal85}. This spectrum was
digitised from the published graph for this paper \footnote{Using
  software available at https://automeris.io/WebPlotDigitizer/}.
Apart from a few differences due to the different resolving power of
the Palomar and Steward spectra, all the circular polarisation spectra
from the 1970s appear essentially identical. When the large
differences in resolving power are taken into account, the polarised
spectra are also in general agreement with the results of filter
polarimetry. Overall, the available data show no convincing evidence
for variability over the period of about six years during the 1970s
when the star was actively observed.

\citet{Putney95} published a circularly polarised spectrum
of \grw\ (see her Fig.~3), with a 11.1\,\AA\ spectral resolution,
in the spectral range 3900-8900\,\AA. The spectrum was presented as
``unpublished data'' from Keck, and used for comparison with the
spectra of other MWDs. We assume that were obtained
around the time of other similar spectra presented in the paper,
i.e., late 1994 or early 1995. Also this spectrum was digitised by
us for its use in this paper.

WHT archive data revealed that the star was observed a number of times
with the ISIS instrument, both in spectroscopic and
spectropolarimetric mode. In
this paper we consider circular spectropolarimetric observations
obtained in 2004-08-05 at UT 21:28, obtained for a total 480\,s
exposure time with grating R600B in the wavelength range 4000 --
5200\,\AA\ and a 1.12\arcsec\ slit width. Setting is very similar to
what we have used in the blue arm, except that, compared to our data,
the archive spectrum is offset by about 450\,\AA\ to the red. A
lower $S/N$ ratio spectrum was obtained the next night and appears
consistent with the one obtained on August 5, within photon-noise
error bars.

\grw\ has been used as a standard star for circular BBP by
\citet{Butetal09}, who report measurements in the {\it
  UBVRI} filters consistent with previous literature (see their
Table~5), e.g.\ 3.6, 4.0 and 4.1\,\% in the $B$, $V$ and $R$ filters,
respectively. We have found in the literature no other published monitoring
of circular BBP of this star.

\subsection{Observations of linear polarisation in the literature}\label{Sect_GrLin}
In addition to a series of circular polarisation measurements 
(Sect.~\ref{Sect_GrCir}), \citet{AngLan70b} discovered strong linear
polarisation, and measured the relevant Stokes parameters using four
different filters in the spectral range between 310 and 700\,nm. They
detected a peak polarisation around 380\,nm of 3.7\,\% (but no linear
polarisation at $\lambda \ga 500$\,nm), with a position angle in the
blue between $16^\circ$ and $24^\circ$. \citet{Angetal72} reported
similar values measured in September 1970 and July 1971.

A single MCSP linear polarisation spectrum, observed with resolutions 
160 and 360\AA\ (in blue and red respectively) in August 1972, 
in the range 329 to 1092\,nm, was reported by
\citet{LanAng75}; they also reported a private
communication from Gehrels suggesting detection of a rotation of the
position angle at 560\,nm.

Another MCSP linear polarisation spectrum, very similar to the one
reported by \citet{LanAng75} was obtained on 28 April 1975 by Angel
and Landstreet (unpublished). Over most of the spectral window
studied, these two Palomar linear polarisation spectra are very similar but
perhaps not quite identical.

A single linear polarisation spectrum, very similar to the circular
polarisation spectrum from the Steward Observatory team described in
the previous section, was obtained in 1976 by \citet{Angetal85}. This
spectrum (digitsed as for the circular polarisation spectrum from the
same paper) shows somewhat more significant changes relative to the two
Palomar linear polarisation spectra. Some of these differences are
certainly due to the markedly higher resolving power of the Steward
spectrum relative to the Palomar spectra in the blue, but other
differences, between 5500--6500\,\AA, may possibly be real. Overall, however,
these data, together with the BBP observations, do not provide
convincing evidence of variability of the linear polarisation during
the 1970s.

\citet{West89} obtained a single BBP measurement during 1986 in the
$B$ filter: 3.19\,\% with $\Theta=19^\circ.76$.

\citet{FriJor01} searched for variability due to rapid rotation of
linear polarisation using a trailing technique at the 2.2\,m telescope
of Calar Alto in the Johnson B filter.  They confirmed that the linear
polarisation of the star is not variable on a time-scale of minutes.

\subsection{New intensity and polarisation spectra}\label{Sect_Our_Observations}
Our new data consist of intermediate resolution flux and polarised
spectra obtained in 2015 and 2017.

In the MWDs with the largest fields, the spectral features of the
intensity spectrum are often shallow, and display low contrast with
respect to the continuum, hence they might not be obvious to the
eye. In an effort to make these features stand out more clearly
without requiring extremely accurate flux calibration of our spectra,
we have experimented the use of the quantity $\delta \mathcal{F}_{\rm
  N}$. This is defined as the difference between the intensity
spectrum as observed ($\mathcal{F}$) and its smoothed version
$\mathcal{F}_{\rm s}$, normalised to the smoothed spectrum:
\begin{equation}
\delta \mathcal{F}_{\rm N} = \frac{\mathcal{F}- \mathcal{F}_{\rm s}}{\mathcal{F}_{\rm s}}
\label{Eq_Smooflux}
\end{equation}
where the smoothed intensity spectrum is calculated with a Fourier
filter $N$\,\AA\ wide.  Obviously, this quantity is prone to enhance
not only stellar features but also spectral features due to the
Earth's atmosphere and spurious instrumental effects (for example
coming from the CCD). However, this function is found to be very
helpful in identifying various spectral features in the intensity
spectrum, therefore we have decided to adopt it.

\begin{figure*}
  \includegraphics*[width=18cm,trim={0.8cm 1.6cm 0.0cm 1.0cm},clip]{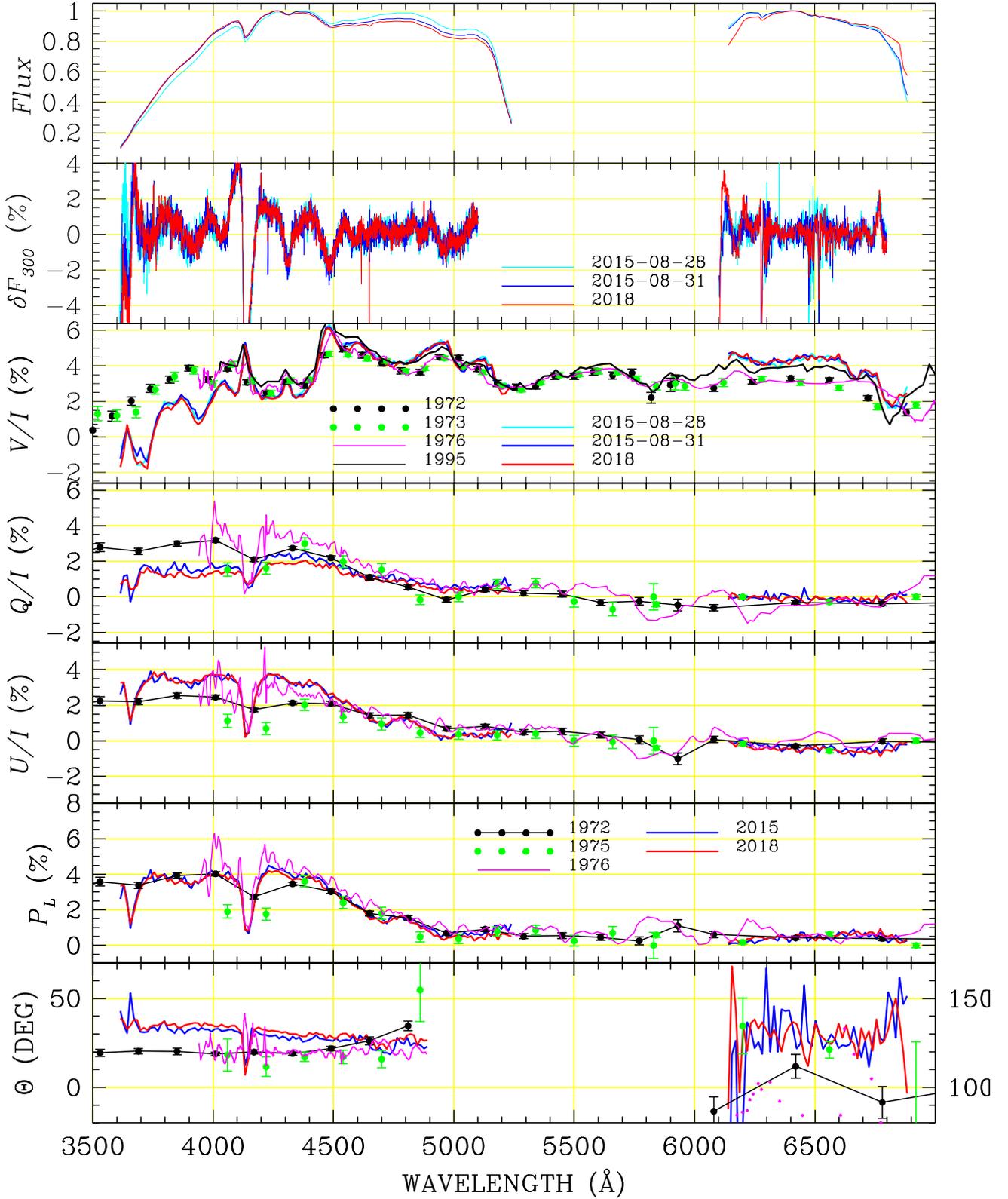}
  \caption{\label{Fig_WD19} Polarisation spectra of \grw\ obtained at
    various epochs. For display purposes, the position angle data $\Theta$
    in the red ($\lambda \ga 6000$\,\AA, see $y$ axis on the right)
    are reported with a 100\degr\ offset with respect to the
    data in the blue ($\lambda \la 5000$\,\AA, see $y$ axis on the left).
    Keys to the symbol are given in the second panel from the top (for flux measurements)
    in the third panel from the top (for $V/I$ measurements) and in the 
    sixth panel from the top (for $Q/I$, $U/I$, \pl, $\Theta$
    measurements). References are given in the text.}
  \end{figure*}

Figure~\ref{Fig_WD19} presents both archival and new polarisation
spectra of \grw. Table~\ref{Table_Grw} gives the observing log and
reports the broadband values of our new polarisation spectra
calculated via Eq.~(\ref{Eq_BBP}).

From top to bottom, the various panels of Fig.~\ref{Fig_WD19} show 1)
the measured (unpolarised) flux $\mathcal{F}$, not corrected for the
instrument transmission function; 2) the difference $\delta
\mathcal{F}_{\rm 300}$ defined by Eq.~(\ref{Eq_Smooflux}) (limited
to our new WHT data only); 3) the reduced Stokes parameter $V/I$; 4)
the reduced Stokes parameter $Q/I$; 5) the reduced Stokes parameter
$U/I$; 6) the fractional linear polarisation \pl; and 7) the
polarisation position angle $\Theta$.  Literature data shown in
Fig.~\ref{Fig_WD19} are those obtained at Mount Palomar in 1972
(circular and linear), 1973 (circular only), and 1975 (linear only),
as well as the Steward spectra (linear and circular) from 1976.
\input{Table_GRW}

The interpretation of the spectrum of \grw\ remained a mystery
for more than a decade after the discovery of its magnetic field.
The main spectral line features were identified when
calculations of the wavelengths of line components of the spectrum of
hydrogen in fields of hundreds of MG were successfully made by
\citet{Roesetal84}. Using these data, \citet{Angetal85} were able to
account qualitatively for all the principal absorption features in the
optical spectrum as being due to H in a magnetic field ranging in
strength between about 160 and 320\,MG. This interpretation of the
observed spectra was confirmed by comparison of the observed Stokes
$I$ spectrum with spectrum synthesis calculations of \citet{Jordan92},
who however showed that the atomic data were not yet sufficient to
make possible modelling of the polarisation spectra. 

In the blue ISIS spectrum, the well-known Minkowski band at
4135\,\AA\ \citep{GreMat57} stands out spectacularly both in
$\delta \mathcal{F}_{\rm N}$ and in Stokes $Q/I$, $U/I$ and $V/I$.
Other Minkowski bands at 3650\,\AA\ and
4466\,\AA\ are also visible.  Additional, somewhat weaker but
apparently real absorption features confirmed by corresponding
polarisation features are found at about 4300\,\AA, 4480\,\AA, and
possibly at 4950\AA. Corresponding absorption lines are seen in some
of the model Stokes $I$ spectra computed by \citet{Jordan92}; because
these are quite weak features in the raw Stokes $I$ spectrum, it is
not clear if most of them are present in the observed $I$ spectrum of
\citet{Angetal85}.  A very weak feature in the Stokes $I$ spectrum at
6650\,\AA\ also correponds to weak but quite clear feature in the
$\delta \mathcal{F}_{\rm N}$, $V/I$ and $Q/I$ spectra.
This feature may correspond to the 3s'0--2p+1
component of H$\alpha$, which is stationary at about 6643\,\AA\ at a
field strength of about 140\,MG and appears in the syntheses of
\citet{Jordan92}.

\begin{table}
  \caption{\label{Tab_Filters} Polarisation of standard star HD\,25443 and of MWD \grw\
    (observed on 2018-09-22) integrated in different $B$ filters as explained in the text.}
  \begin{center}
  \begin{tabular}{lrr|rrr}
    \hline\hline
              & \multicolumn{2}{c}{HD\,25443} & \multicolumn{3}{c}{\grw}   \\
              &   \pl (\%) &$\Theta$ (\degr)  & \pl (\%) &$\Theta$ (\degr) & \pv (\%) \\
\hline
FORS B Bessel & 5.11  & 138.1                 & 3.08 & 31.5 & 3.22\\
FORS b\_high  & 5.14  & 138.1                 & 2.76 & 30.7 & 3.58 \\
ACAM B Bessel & 5.13  & 138.1                 & 2.86 & 31.0 & 3.43 \\
\hline
  \end{tabular}
  \end{center}
\end{table}

\section{Broadband polarimetry measurements may not be suitable
to detect real variability}
Since this work is primarily devoted to the search for subtle variablity
of \grw\, we make a preliminary comment about BBP
measurements.

In the blue spectral region, the polarisation of \grw\ changes rapidly
with wavelength. Therefore, BBP measurements obtained with slightly
different instruments settings may differ one from each other, even if
the polarisation intrinsic to the star is constant. This can be easily
seen via numerical simulations through Eq.~(\ref{Eq_BBP}), using the
observed spectra but different filter transmission curves. As an
example, Table~\ref{Tab_Filters} shows the integrated polarisation
calculated using Eq.~(\ref{Eq_BBP}) and the transmission functions of
three different broadband $B$ filters: an old $B$ Bessel filter used
with the FORS instrument of the ESO VLT, the currently used FORS
b\_high filter, and the $B$ Bessel filter used with the ACAM
instrument of the WHT (filter transmission curves may be found in the
respective instrument web pages). For \grw, the results differ by
0.2--0.3\,\%, which may well be larger than
uncertainties due to photon noise, and hence easily detected.

This excercise shows that subtle changes observed in BBP
measurements may actually be due to small differences of the
transmission function of the instrument+telescope, rather than stellar
variability. It is important to note that such differences due to
different filters would not be
detected in the observations of a standard star for linear polarisation.
Table~\ref{Tab_Filters} shows that, due to its smooth behaviour with
wavelength, the differences in the wavelength integrated polarisation
of a standard star are of the order of $10^{-4}$, i.e., some 10 times
smaller than those calculated for \grw. Even using the same instrument
and instrument setup for broad band measurement may not lead to conclusive results because
instrument sensitivity may change over the years in a way not detectable
through the monitoring of standard stars.

\section{Has the polarised spectrum of \grw\ changed with time?}\label{Sect_Change}
Inspection of the data from the 1970s in Fig.~\ref{Fig_WD19} confirms
the conclusions of the original literature, i.e., that circular
polarisation spectra obtained from 1972 to 1976 appear very similar to
each other, and that the linear polarisation spectra obtained over
the same period of time show at most some small differences in the
blue spectral regions.
Our new measurements obtained in 2015 and 2018 also seem {\it nearly}
identical to each other.

Compared to the measurements in the 1970s, our new circular
polarisation clearly show significant differences both in the red and
at the shortest wavelengths. Linear polarisation shows obvious changes
in the blue, and perhaps marginal differences in the red. Since about
40 years have elapsed between the observations of
\citet{Angetal72,LanAng75,Angetal85} and our observations,
polarisation measurements set a time scale for the changes of the
order of decades or more.

In the following we inspect and comment on these differences in more
detail.

\subsection{Intensity spectra}
ISIS data reveal that $\delta \mathcal{F}_{\rm N}$ is constant within
the 3-year interval of time from 2015 to 2018, both in the blue and in
the red part of the observed spectrum. Unfortunately, we are not able to perform a
meaningful comparison with intensity data obtained in previous
decades.

\subsection{Circular polarisation}
As noted above, circular polarisation spectra obtained between 1972 and
1976 appear very similar to each other, as do the three
circular polarisation spectra obtained by us in 2015 and
2018. However, when comparing datasets obtained 40 years apart, the
amplitude of cirular polarisation has changed significantly at
$\lambda \la 4000$\,\AA, betwen 4400 and 5000\,\AA\ and between 6100
and 6700\,\AA. Spectra obtained by \citet{Putney95} and WHT archive
data obtained in 2004 show also some differences as discussed below.

\subsubsection{The blue}
\begin{figure}
  \includegraphics*[width=8cm,trim={0.5cm 6.2cm 1.5cm 3.5cm},clip]{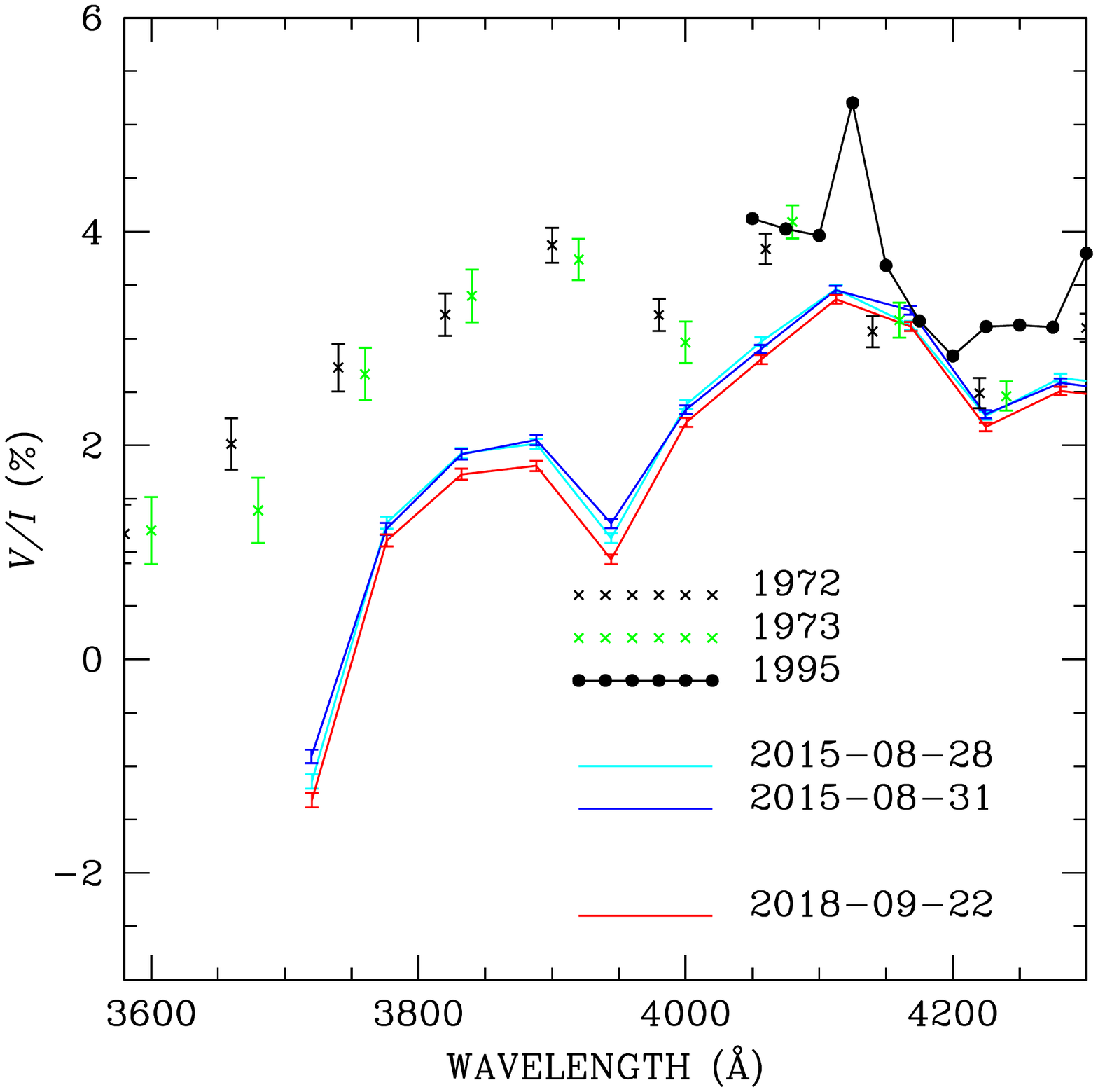}
  \includegraphics*[width=8cm,trim={0.5cm 6.2cm 1.5cm 3.5cm},clip]{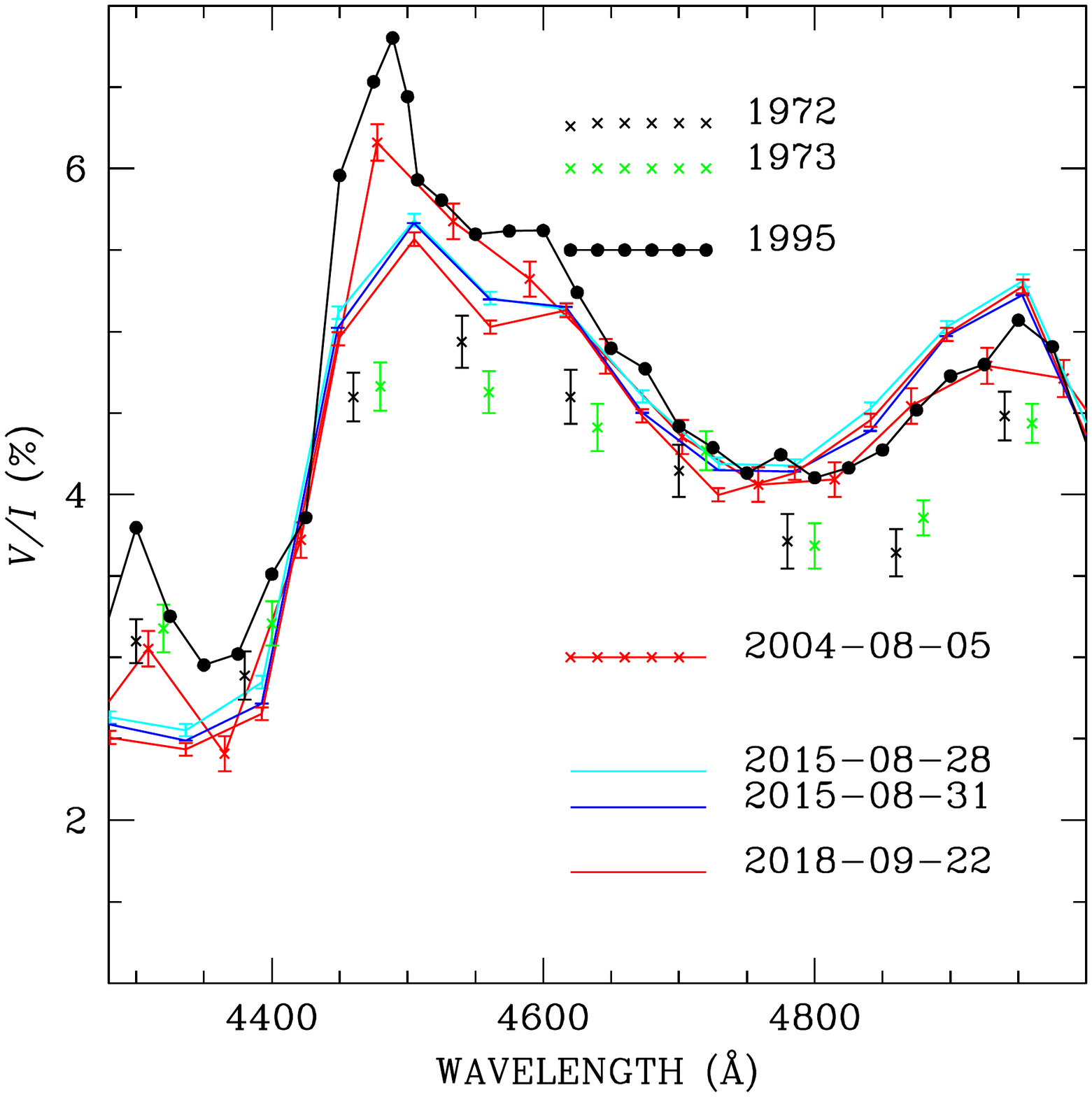}
  \includegraphics*[width=8cm,trim={0.5cm 5.3cm 1.5cm 3.5cm},clip]{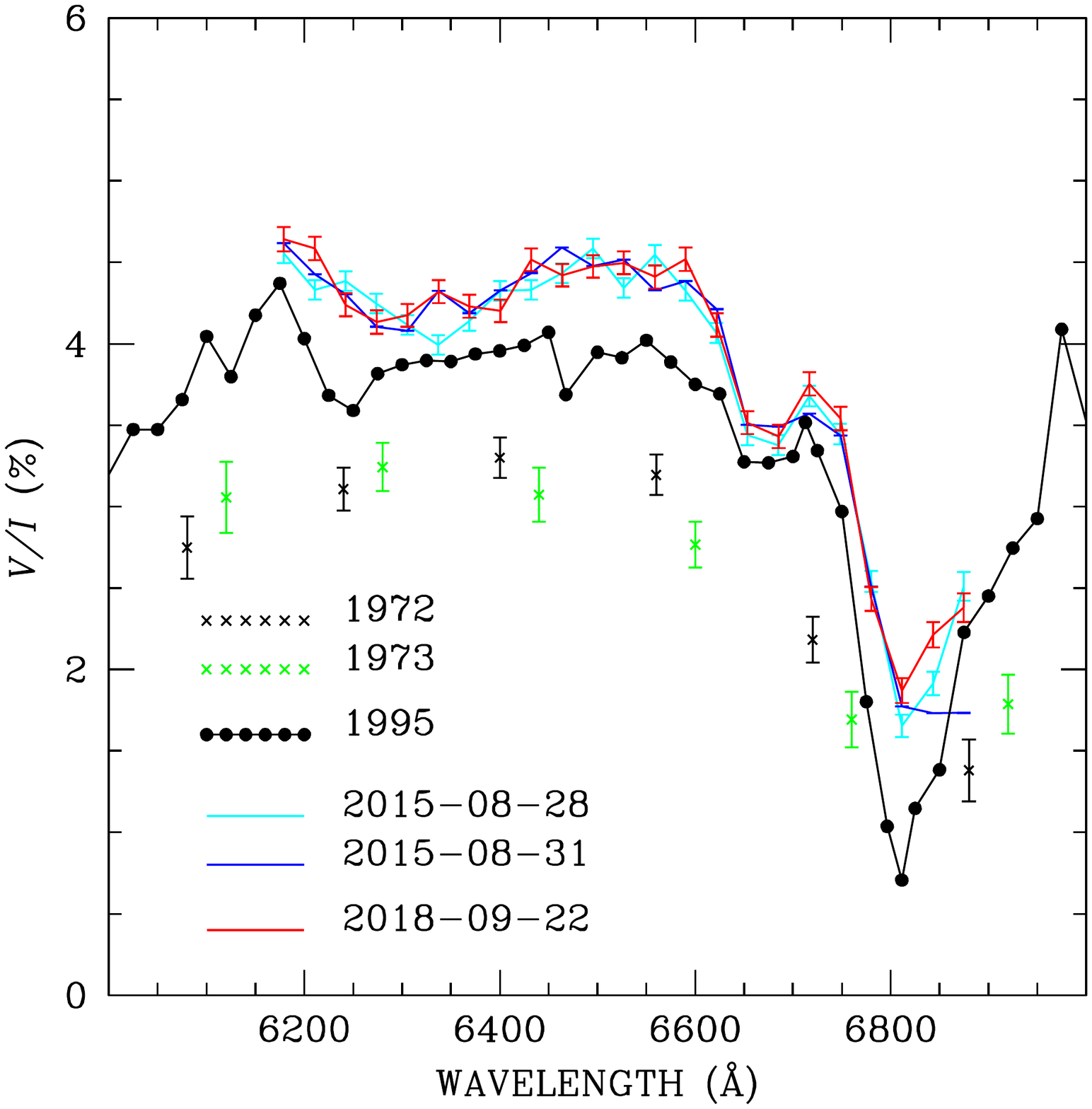}
  \caption{\label{Fig_BlueVis} Enlarged view of circular spectropolarimetric
    data in spectral region covered by the ISIS instrument where discrepancies
    among datasets are noted.}
  \end{figure}
The most remarkable variation in the circular polarisation spectra of
\grw\ is seen at $\lambda \la 4000$\,\AA. Compared to the spectra
obtained in the 1970s, circular polarisation is offset by about
$-1$\,\%. A careful inspection also shows that while the two datasets
obtained in 2015 are fully consistent among themselves (within error
bars), the polarisation measured in 2018 is slightly different
precisely in those wavelength regions where large differences are
detected when we compare data taken 40-45 years apart. This is best
seen in Fig.~\ref{Fig_BlueVis}. Differences bewteen 2015 and 2018
datasets are tiny (e.g. $\sim-0.2$\,\% at $\lambda\la
4000$\,\AA). They are above the uncertainty due to photon noise, but
not large enough to exclude that they are entirely due to some
systematics. However, the fact that they appear in the same spectral
region where large changes are seen in a 40-years interval of time
suggests that the variability detected during a three-year span may
well be intrinsic to the source (rather than due to an instrumental
effect). Obviously this conclusion should be confirmed by further
monitoring.

Differences are prominent also betweeen 4400 and 4600\,\AA, where
circular polarisation had a maximum in 1995, then decreased in 2004
and continued to decrease in 2015 and 2018.

Due to the large variations of the polarisation in the Minkowski
bands, a comparison between our measurements synthetically integrated
in the blue (3.43\,\% in a $B$ Bessel filter) and the
$3.61\pm0.11$\,\% value measured in the $B$ filter by
\citet{Butetal09} is not meaningful, apart from confirming that the
polarisation does not change much within a time-scale of a few years.

\subsubsection{The red}
Circular polarisation around 6500\,\AA\ seems to have increase by
about 1.5\,\% from the 1970s (when it was about 3\,\%) to $\sim 4.5$\,\% at
the time of our observations. In this spectral region we have not
detected any change between 2015 and 2018, but data obtained in 1995
were in between the values measured in the two epochs. At 6800\,\AA,
data obtained by \citet{Putney95} show the lowest values among
all datasets ($\sim 1$\,\%).

Our BBP measurements
in the red (4.0\,\% integrated over a $R$ filter) are consistent with
the relatively recent BBP measurement of \citet{Butetal09}, who had
detected a signal of $4.06\pm0.16$\,\%.  In conclusion, we have
clearly detected a change of 1.5\,\% over a 40+ years interval, but we have
found no evidence of change over a time-scale of the order of three
years.

\subsection{Linear polarisation}
No large changes of the fraction of linear polarisation are detected
either on a short (3-year) or long (40 year) time-scale. However, over a 40--45 year
interval of time, the position angle has definitely changed by more than
10\degr\ both in the blue and in the red. In the following we look
at the various linear polarisation spectra and their differences in more
detail.

On the long time scale sampled by the difference between the 1970s
data and our recent ISIS spectra (40 years), there definitely are
changes. In particular, below 4000\,\AA, the recent level of linear
polarisation sampled by $Q/I$ seems to have decreased, while that
sampled by $U/I$ is larger. This corresponds to a very significant
rotation of the position angle of the linear polarisation away from
the value of $\sim 20\degr$ found in the 1970's to about 33\degr\ in
recent times.

We note that the measurement by \citet{West89} obtained 1986 suggests
also a decrease of linear polarisation compared to the 1970s (3.19\,\%
in the $B$ filter, for a position angle of $ 19.1 \pm 0.8$) but as
discussed in Sect.~\ref{Sect_BBP} these kinds of comparison should be
made with caution.

It is notable that although the circular polarisation is quite
different from zero throughout the visible spectrum, the non-zero
linear polarisation is present essentially below 5500\,\AA, with
possible low-level features suggested by the Steward spectrum, but
largely absent from the lower resolution MCSP data. The position angle
in the red as measured in the 1970s was around 110\degr, with a fair
amount of scatter due to the small but not quite vanishing amplitude
of $Q$ and $U$ in that region. In the ISIS data, both $Q$ and $U$ are
also close to zero, therefore the position angle has a large uncertainty,
but is  consistent with a 10--30\degr\ rotation from the value measured in
the 1970s.

We conclude that linear polarisation measurements are certainly
variable over a time-scale of decades, mostly in terms of a rotation,
while no significant variation has been detected on time scale of
three years. A rotation constant in wavelength provides a first rough description
of the change in linear polarisation.

\section{Constraints on the rotation period of \grw}
\begin{figure}
  \includegraphics*[width=4.3cm,trim={2.5cm 5.1cm 1.5cm 2.0cm},clip]{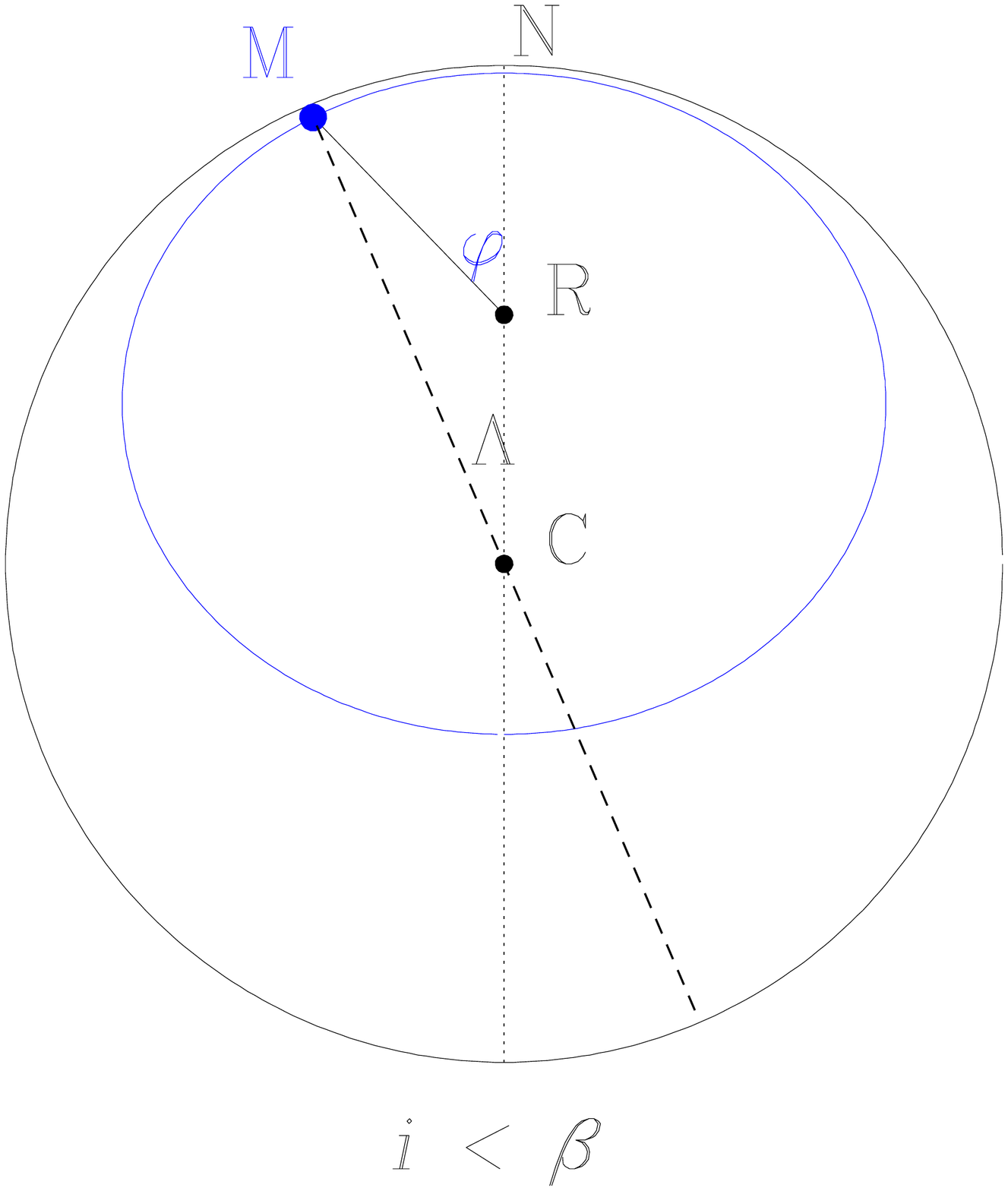}\ 
  \includegraphics*[width=4.3cm,trim={2.5cm 5.1cm 1.5cm 2.0cm},clip]{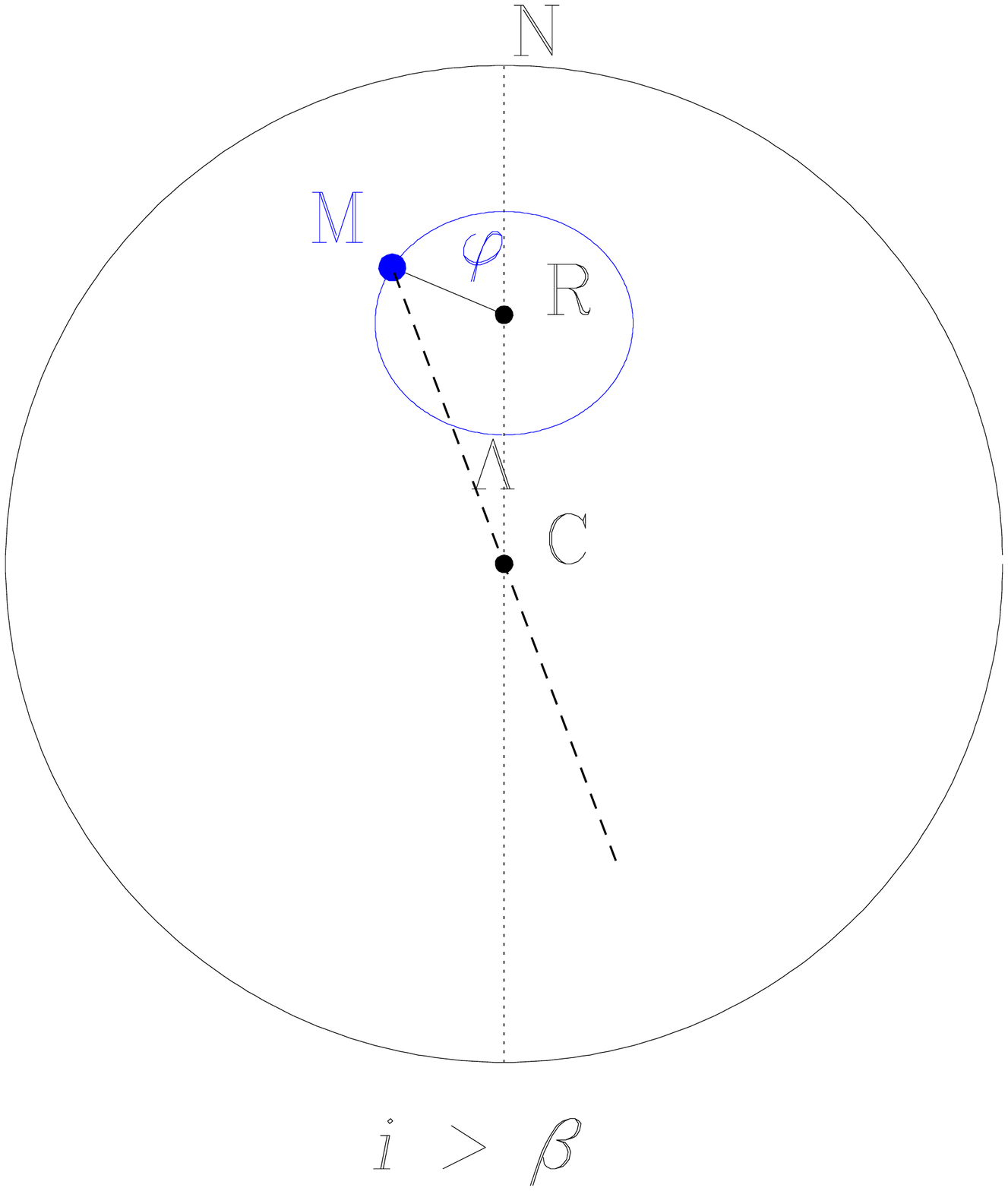}
  \caption{\label{Fig_ORM} Two examples of oblique rotator model. The black solid lines represent the stellar disk;
    the centre of the stellar disc $C$,
    the rotation pole $R$ and the North Celestial Pole $N$ are aligned. $M$ represents the magnetic pole at the surface of the
    star, and the blue solid line represents its trajectory on the stellar disk as the star rotates ($\varphi$ represents the rotation angle).
    The dashed line represents the projection of the dipolar axis in
    the plane of the sky, and forms and angle $\Lambda$ with the direction to the North Celestial Pole.
    {\it Left panel:}  $\beta > i$ (as the star rotates, the position angle $\Lambda$ spans monotonically the range 0-360\degr,
    and the polarisation position angle spans twice the range 0-180\degr).
    {\it Right panel:} $\beta < i$ (as the star rotates, the position angle $\Lambda$ spans
    a limited angle range, and changes direction twice, and so does the polarisation position angle). }
  \end{figure}
Measurements of the position angle of the broadband linear
polarisation of \grw\ in the blue between about 3800 and
4600\,\AA\ may be used to try to detect evidence of stellar
rotation, under the assumption that the polarisation position angle
$\Theta$ tracks the position angle of the magnetic axis $\Lambda$.

We assume that the star may be described by an oblique rotator model,
with rotation axis tilted at an angle $i$ with respect to the line of
sight, and that the magnetic axis, for example a dipole axis, is tilted at an
angle $\beta$ with respect to the rotation axis.  Formally, the angle
$i$ identifies the positive rotation pole, and the angle $\beta$
identifies the positive magnetic pole, and both $i$ and $\beta$ may
range from 0\degr\ to 180\degr \citep[e.g.][]{Lanetal98}. For
simplicity, we will assume that $i \le 90\degr$ and $\beta \le
90\degr$, a situation to which one can always be brought back to, after
reversing the rotation direction, the polarity of the magnetic
field, or both \citep[see Sect.~3 of][]{Lanetal98}.  None of these
transformations will affect our conclusions below, which depend
neither on the direction of the rotation, nor on the sign of the
magnetic field.

There are two rather different geometrical situations that we should
consider when interpreting time variations in the observed position
angle of linear polarisation. One situation is when the tilt angle of
the rotation axis $i$ is smaller than the obliquity of the magnetic
axis $\beta$, and one when the tilt angle $i$ is larger than the
obliquity $\beta$.

(1) When $i < \beta$, the position angle of the dipolar axis $\Lambda$
will span the range 0 to 360\degr\ as the star rotates.  This
situation is visualised in the left panel of Fig.~\ref{Fig_ORM}, and
may be more easily understood by thinking of a star with the rotation
axis parallel to the line of sight ($i=0\degr$) and the magnetic axis
perpendicular to the rotation axis ($\beta = 90\degr$).  As the star
rotates, the magnetic pole completes a 360\degr\ tour in the plane of
the sky, and the polarisation spans twice the range $0-180$\degr,
always rotating in the same direction. Note, however, that the
specific example of a rotation axis perpendicular to the line of sight
implies that the longitudinal component of the magnetic field should
be constant as the star rotates; this is not consistent with the
observations of circular polarisation of \grw, which show variability
with time.

(2) When $i > \beta$, the projected dipolar axis will not complete a
full 360\degr\ as the star rotates, but would simply oscillate on the
sky, changing direction every half rotation period (see the right
panel of Fig.~\ref{Fig_ORM}).  The amplitude of the wobble of the
position angle would be larger, the larger the value of $\beta$.

Because the position angle of linear polarisation may be quite
precisely determined, searching for changes in the direction of the
position angle is clearly a relatively straight-forward method of
detecting stellar rotation, unless the star rotates extremely slowly,
or the field axis and the rotation axis are very closely parallel.

As we have detected rotation of the polarisation position angle, which
we assume is due to rotation of the underlying star rather than
significant evolution of the stellar magnetic field structure, it is
of interest to try to obtain constraints on the stellar rotation
period from the observations. Of course, we are severely limited by
the obvious fact that we have not yet observed one full rotation, and
so we do not yet know whether the geometry of the rotation axis and
field obliquity is closer to case (1) or (2).

If the situation is close to case (1), with $i \le \beta$, then a
single rotation produces two full circles of the position angle on the
sky. In this case we might have observed 10--15$^\circ$ of a full
360$^\circ$ rotation in about 40 years, suggesting a rotation period
of the order of one millenium.\footnote{It was this argument, applied
  in the 1970s to an upper limit of perhaps $4^\circ$ of position
  angle rotation in two years of observation, that led to the
  suggestion of a period of the order of 100\,yr or longer
  \citep{LanAng75}, which has frequently been repeated in the
  literature.}

If \grw\ is closer to case (2), the minimum possible period would
correspond to our two main epochs of observation corresponding roughly
to the two extremes of position angle oscillation. In this case
another 40 or 50 years might bring the position angle back to about
20$^\circ$. If this is the situation, the rotation period might be as
short as a century.

In this discussion, we have made the implicit assumption that the
polarisation position angle has monotonically changed during the last
40 years. In fact, we are not aware of linear polarisation measurements that can validate
this. Also, we note that our period limit estimates are based on the
assumption that ${\rm d} \Lambda / {\rm d}t$ is constant with
time. This clearly cannot be true, around the epoch of inversion of
the direction of changes of $\Lambda$, in scenario no.\ 2. Also, the effect
of Faraday rotation may change as the star rotates, affecting also
the assumption that  ${\rm d} \Lambda / {\rm d}t$ is constant with time.
However, our simple modelling should be sufficient for an order-of-magnitude
estimate of the rotation period, which is all we can extract from the
very limited available data.  We conclude that the rotation period
lies probably in the range of 10$^2$ to 10$^3$ yr.

Our period estimate may seem outrageously long, but there are main
sequence stars known to have rotation periods near the lower limit,
for example the star HD\,201601 = $\gamma$\,Equ, with a period
estimated to be about one century \citep{Leroetal94,Bychetal16}. It is
clear that physical processes exist that are capable of removing
virtually all of a star's angular momentum, and that one of the most powerful of these
processes acts through the stellar magnetic field.

\section{Conclusions}
We have presented new polarisation spectra of the white dwarf \grw,
one of the most strongly magnetic of the known MWDs, and searched for
variability.

To carry out our investigation, we had to perform a careful analsysis
of various polarimetric measurements obtained with the ISIS
instruments in the last few years. We found some important results
that help to characterise the instrument, and that may be useful to other
users interested in the polarimetric capabilities at the WHT. In particular
we found that:

\begin{itemize}
\item Both linear and circular instrumental polarisation are smaller
  (in absolute value) than 0.05\,\%.
\item Cross-talk from linear to circular polarisation is negligible
  for most practical situations. Our estimate is that less than
  1\,\% of the linear polarisation signal is transformed into
  circular polarisation.
\item We found no evidence that the use of a dichroic beam splitter introduces a
  spurious linear polarisation signal, confirming what was already found
  for circular polarisation by \citet{BagLan18}.
\item We have presented a list of ISIS observations of standard stars
  for linear polarisation to which we may refer in the future to
  monitor instrument stability,
\item We have used lunar spectra, twilight spectra and background
  spectra to confirm the alignment of the polarimetric optics. This
  method may be adopted more systematically as it is probably faster
  and more accurate than observing standard stars.
\item We have found inconsistencies between the way the circular
  polarisation of \grw\ has been reported over the years; we have
  discussed the definition of the sign of circular polarisation, and
  clarified what we have adopted.
\end{itemize}

Regarding the scientific investigation of \grw:
\begin{itemize}

\item We have shown that spectropolarimetry in all four Stokes
  parameters (as opposed to broad band polarization measurement) is the
  method of choice for the study of strongly magnetic white dwarfs for
  two reasons. One is that only spectropolarimetry may reveal the
  shape of the Stokes profiles of the spectral features; the other 
  is that it allows to discriminate between changes of the instrument
  sensitivity and changes of the polarisation intrinsic to the
  source.

\item We have clearly established the variability of the polarisation
  spectra of the MWD \grw. We have argued that the observed rotation
  of the position angle of linear polarisation suggests that
  the rotation period is probably in the range of 10$^2$ to
  10$^3$\,yr. This result would imply that \grw\ is  one of
  the most slowly rotating stars known.

\end{itemize}

At this early stage in re-starting systematic observations of the
highest-field MWDs, we hope that we have re-launched the systematic
observation of these mysterious and potentially very important
objects, which contain the largest fields in the universe that are
easily observable directly in optical light. One of the long-term
goals of this work is to obtain the observations needed to determine
the rotation period of \grw, and to map the surface field structure, a
task that we are obliged to leave to future generations of
astronomers.

\section*{Acknowledgments}

This work is based on observations collected at the William Herschel
Telescope, operated on the island of La Palma by the Isaac Newton
Group, programmes P17 during semester 15B, and programme P15 during
semested 18B. Important ISIS calibration data were obtained in the
context of programme P5 during semester 14A, programme P30 during
semester 14B, P28 during semester 15A. This paper has also made use
of data obtained from the Isaac Newton Group Archive which is maintained
as part of the CASU Astronomical Data Centre at the Institute of Astronomy,
Cambridge. JDL acknowledges the financial support of the Natural Sciences and
Engineering Research Council of Canada (NSERC), funding reference
number 6377-2016. We thank Ian Skillen (ING) for 
calculating the instrument position angle on sky when
the instrument and telescope were in the parking position,
Tom Marsh for 2004 WHT data of \grw\ and Stefan Jordan for his useful review
of this manuscript.

\bibliographystyle{mnras}
\bibliography{sbabib}
\label{lastpage}
\end{document}

%% file: Table_ZeroPol.tex
\begin{table*}
  \caption{\label{Table_ZeroPol} Observations of zero polarised stars.}
  \begin{center}
  \begin{tabular}{lccrrlr@{$\pm$}lr@{$\pm$}lr@{$\pm$}lr@{$\pm$}lr}
    \hline
    \hline
          &    DATE    & UT  &  EXP &    PA    &         &
    \multicolumn{2}{c}{\pq\ (\%)} &
    \multicolumn{2}{c}{\pu\ (\%)} &
    \multicolumn{2}{c}{\pq\ (\%)} &
    \multicolumn{2}{c}{\pu\ (\%)} & \\
    STAR    & yyyy-mm-dd &hh:mm&  (s) &($^\circ$) & setting &
    \multicolumn{4}{c}{$B$}&\multicolumn{4}{c}{$R$}& Ref.\\
  \hline
HD\,144287 &  2014-03-07 &05:00& 100 & 285 & R600R       &\multicolumn{4}{c}{-- }     &$ 0.01$& 0.01&$ 0.00$&  0.00&1\\
           &  2014-03-07 &05:05& 100 & 285 & R600B       &$ 0.00$& 0.02&$-0.01$& 0.02 &\multicolumn{4}{c}{-- }     & \\ 
           &  2014-03-07 &05:09& 100 & 285 & R600B+R600R &$-0.02$& 0.03&$ 0.02$& 0.03 &$-0.02$& 0.00&$ 0.00$&  0.00& \\[1mm]
           &  2014-03-07 &05:28& 120 & 195 & R600R       &\multicolumn{4}{c}{-- }     &$-0.00$& 0.00&$-0.01$&  0.00& \\
           &  2014-03-07 &05:33& 120 & 195 & R600B       &$-0.03$& 0.00&$-0.02$& 0.00 &\multicolumn{4}{c}{-- }     & \\
           &  2014-03-07 &05:38& 120 & 195 & R600B+R600R &$-0.03$& 0.02&$-0.01$& 0.02 &$-0.01$& 0.02&$-0.01$&  0.02& \\[2mm]
HD\,98161  &  2015-01-06 &05:33& 240 &   4 & R300B       &$ 0.02$& 0.09&$-0.08$& 0.09 &\multicolumn{4}{c}{-- }     &2\\
           &  2015-01-06 &05:50& 120 &   4 & R158R       &\multicolumn{4}{c}{-- }     &$ 0.02$& 0.01&$ 0.00$& 0.01 & \\[2mm]
HD\,125184 &  2015-01-06 &06:21&  40 & 321 & R158R       &\multicolumn{4}{c}{-- }     &$ 0.01$& 0.02&$-0.01$&  0.01&1 \\
           &  2015-01-06 &06:24&  40 & 321 & R300B       &$-0.05$& 0.01&$ 0.02$& 0.01 &\multicolumn{4}{c}{-- }     & \\[2mm]
HD\,28099  &  2018-09-24 &06:16&  35 & 240 & R600B+1200R &$-0.03$& 0.00&$-0.02$& 0.00 &$-0.01$& 0.00&$-0.04$& 0.00 &3\\[2mm]
\hline
  \end{tabular}
\end{center}
  \noindent
  Key to references: 
  {\bf 1.} \citet{Serkowski74}\ \ {\bf 2.} \citet{Turetal90}\ \ {\bf 3.} Not previously used as unpolarised reference
  star -- see text.
\end{table*}

%% file: Table_STD.tex
\begin{table*}
  \caption{\label{Table_STD} Observations of standard stars for linear polarisation. Position angle is measured counterclockwise
    looking at the source from the great circle passing through the star and the North Celestial Pole. For comparison, literature
    data are reported in boldface fonts. Uncertainties are due to photon-noise, and do not take into account systematics.
    Observations marked with asterisks are likely to be affected by the contamination of a strong background for the reasons
    explained in the notes.
  }
  \tabcolsep=0.12cm
  \begin{tabular}{lcrrlr@{$\pm$}lr@{$\pm$}lrr@{$\pm$}lr@{$\pm$}lr}
    \hline
    \hline
          &    DATE    & UT  &  EXP &                 &
    \multicolumn{2}{c}{\pl\ (\%)} &\multicolumn{2}{c}{$\Theta$ ($^\circ$)}  & \multicolumn{1}{c}{$\Delta \Theta$ ($^\circ$)} &
    \multicolumn{2}{c}{\pl\ (\%)} &\multicolumn{2}{c}{$\Theta$ ($^\circ$)}  & \multicolumn{1}{c}{$\Delta \Theta$ ($^\circ$)} \\
  STAR    & yyyy-mm-dd &hh:mm&  (s) & setting          &
  \multicolumn{5}{c}{$B$} &  \multicolumn{5}{c}{$R$}\\
  \hline
HD\,161056   & \multicolumn{4}{l}{\citet{Schetal92}}      &{\bf 3.80}&{\bf 0.60}&{\bf 66.6}&{\bf 0.4}&      &{\bf 4.01}&{\bf 0.03}&{\bf 67.3}&{\bf 0.2}&      \\
             &  2014-03-06 & 06:00 &160 & R600R           &\multicolumn{5}{c}{}                             &     3.99 &     0.01 &     68.0 &     0.1 &$+0.7$\\
             &  2014-03-06 & 06:06 & 80 & R600B           &     3.65 &     0.01 &     68.0 &     0.1 &$+1.4$&\multicolumn{5}{c}{}                             \\
             &  2014-03-06 & 06:10 & 80 & R600B+R600R     &     3.65 &     0.03 &     67.9 &     0.3 &$+1.2$&     4.00 &     0.01 &     68.1 &     0.1 &$+0.8$\\ [2mm]
HD\,204827   &\multicolumn{4}{l}{\citet{Schetal92}}       &{\bf 5.65}&{\bf 0.02}&{\bf 58.2}&{\bf 0.1}&      &{\bf 4.89}&{\bf 0.03}&{\bf 59.1}&{\bf 0.2}&      \\
             &  2015-01-05 & 19:24 &  80 & R158R          &\multicolumn{5}{c}{}                             &      4.93 &     0.02&     59.7 &     0.1 &$+0.6$\\
             &  2015-01-05 & 19:32 & 240 & R300B          &     5.73 &     0.00 &     58.6 &     0.1 &$+0.4$& \multicolumn{5}{c}{}                            \\  [2mm]
HD\,154445   & \multicolumn{4}{l}{\citet{Schetal92}}      &{\bf 3.45}&{\bf 0.05}&{\bf 88.9}&{\bf 0.4}&      &{\bf 3.68}&{\bf 0.07}&{\bf 88.9}&{\bf 0.6} &     \\
             &  2015-02-04 & 06:35 & 160& 600B            &     3.44 &      0.04&     89.5 &     0.3 &$+0.6$& \multicolumn{5}{c}{}                            \\ [2mm]
HD\,160529   & \multicolumn{4}{l}{\citet{HsuBre82}}       &{\bf 6.97}&{\bf 0.03}&{\bf 20.1}&{\bf 0.1}&      &{\bf 7.04}&{\bf 0.01}&{\bf 20.8}&{\bf 0.1}&      \\
             &  2015-08-30 & 20:28 &  80 &R1200R          &\multicolumn{5}{c}{}                             &     6.97 &     0.00 &     19.7 &     0.0 &$-1.1$\\ 
             &  2015-08-30 & 20:35 & 240 &R600B           &     6.97 &     0.02 &     19.1 &     0.1 &$-1.0$& \multicolumn{5}{c}{}      \\
             &  2015-08-30 & 20:42 & 120 &R600B+R1200R    &     6.94 &     0.00 &     19.2 &     0.0 &$-0.9$&     6.93 &     0.00 &     19.7 &     0.0 &$-1.1$\\
             &  2015-08-30 & 20:48 & 240 &R1200R          &\multicolumn{5}{c}{}                             &     6.97 &     0.00 &     19.4 &     0.1 &$-1.4$\\ [2mm]
HD\,25443    &\multicolumn{4}{l}{\citet{Schetal92}}       &{\bf 5.23}&{\bf 0.09}&{\bf 134.3}&{\bf 0.5}&      &{\bf 4.73}&{\bf 0.05}&{\bf 133.7}&{\bf 0.3} &   \\
             &  2014-03-11 & 20:31 & 480 &300B            &     5.12 &     0.02 &     135.9 &     0.1 &$+1.6$&\multicolumn{5}{c}{}      \\
             &  2014-03-11 & 21:03 &  80 &158R            &\multicolumn{5}{c}{}                              &     4.80 &     0.01 &     135.3&      0.0&$+1.6$\\
             &  2015-01-06 & 00:17 &  80 &R158R           &\multicolumn{5}{c}{}                              &     4.76 &     0.04 &     135.4&      0.2&$+1.7$\\ 
             &  2015-01-06 & 00:24 & 320 &R300B           &     5.11 &     0.02 &     135.7 &     1.4 &      &\multicolumn{5}{c}{}                             \\
   (*)       &  2015-08-31 & 06:28 & 120 &R1200R          &\multicolumn{5}{c}{}                              &     4.91 &     0.00 &     136.6&      0.0&$+2.9$\\
   (*)       &  2015-08-31 & 06:34 & 120 &R600B           &     5.16 &     0.02 &     136.0 &     0.1 &$+1.7$& \multicolumn{5}{c}{}                             \\ 
   (*)       &  2015-08-31 & 06:38 & 100 &R600B+1200R     &     5.03 &     0.00 &     135.1 &     0.0 &$+0.8$&     5.11 &     0.00 &     135.5&      0.0&$+1.8$\\ 
   (**)      &  2018-09-22 & 03:57 & 480 &R600B+1200R     &     5.11 &     0.01 &     138.1 &     0.1 &$+3.8$&     4.87 &     0.02 &     137.9&      0.1 &$+3.2$ \\ [2mm]
HD\,19820    &\multicolumn{4}{l}{\citet{Schetal92}}       &{\bf 4.70}&{\bf 0.04}&{\bf 115.7}&{\bf 0.2}&      &{\bf 4.53}&{\bf 0.03}&{\bf 114.5}&{\bf 0.2}&      \\
   (**)      &  2018-09-22 & 03:42 & 480 &    R600B+R1200R&     4.59 &     0.02 &     118.4 &     0.1 &$+2.7$&     4.55 &     0.02 &     117.0&      0.1 &$+2.5$\\[2mm]
BD$+$59\,389 & \multicolumn{4}{l}{\citet{Schetal92}}      &{\bf 6.35}&{\bf 0.04}&{\bf  98.1}&{\bf 0.2}&      &{\bf 6.43}&{\bf 0.02}&{\bf  98.1}&{\bf 0.1}       \\
   (**)      &  2018-09-22 & 03:25 & 720 &    R600B+R1200R&     6.20 &     0.03 &     102.0 &     0.2 &$+3.9$&     6.48 &     0.02 &     101.1 &     0.1 &$+2.0$\\ [2mm]
\hline
  \end{tabular}
  \\
  (*) Observations were obtained close to sunrise; sky background was high and rapidly increasing from one exposure to the next.\\
 (**) Observations were taken with a wrong combination of dekker and window readout, therefore sky background could not be measured and subtracted.
\end{table*}
\begin{table}
  \caption{\label{Table_CALIBS} Calibrations of the polarisation position angle in the blue arm}
    \tabcolsep=0.12cm
\begin{small}
    \begin{tabular}{lcrr@{$\pm$}lr}
    \hline
    \hline
          &    DATE\ \ \  UT  &
    \multicolumn{3}{c}{$\Theta$ ($^\circ$)}  & \multicolumn{1}{c}{$\Delta \Theta$ ($^\circ$)}\\
OBJECT    & yyyy-mm-dd hh:mm&Expected&\multicolumn{2}{c}{Observed}& \\

  \hline

Moon         &  2015-02-01 20:16 &  175.8 & 175.7 &  1.1 &$-0.1$ \\
             &  2015-09-01 06:06 &  159.0 & 159.0 &  0.4 &$ 0.0$ \\
             &  2018-09-22 19:38 &  151.8 & 151.7 &  0.2 &$-0.1$ \\ [2mm]
Twilight     &  2018-09-22 19:38 &    3.8 &   5.1 &  0.0 &$+1.3$ \\
             &  2018-09-23 19:09 &    0.6 & 179.7 &  0.1 &$-0.9$ \\ [2mm]
Background   &  2015-09-01 00:17 &    3.9 &   1.7 &  0.3 &$-2.2$ \\
             &  2015-09-22 21:01 &   39.2 &  39.6 &  0.3 &$+0.4$ \\
\hline
    \end{tabular}
    \end{small}
\end{table}

%% file: Table_GRW.tex
\begin{table*}
  \caption{\label{Table_Grw} Log of our new observations of \grw. We show also the BBP values obtained using the transmission curves of the old
    FORS B Bessel filter (cols.\ 3-7) and of the FORS Red Bessel filter (cols.\ 8-12). As in Table~\ref{Table_STD}, uncertainties
    reflect photon-noise. Systematics errors are probably of the order of 0.1-0.2\,\% at most for \pv, \pq, \pu, and \pl, and 1--2\degr\ for the position
    angle $\Theta$. The large uncertainties of $\Theta$ in the red are due to the fact that linear polarisation is very small, hence
    its position angle not well determined. For the reasons explained in the text, these values are only internally
    self-consistent, and should not be compared to BBP measurements obtained with other instrument/telescopes.}
  \begin{small}
    \tabcolsep=0.14cm
  \begin{tabular}{ccr@{$\pm$}lr@{$\pm$}lr@{$\pm$}lr@{$\pm$}lr@{$\pm$}lr@{$\pm$}lr@{$\pm$}lr@{$\pm$}lr@{$\pm$}lr@{$\pm$}l}
  \hline
  \hline
           &        &       
\multicolumn{10}{c}{$B$}&
\multicolumn{10}{c}{$R$}\\
DATE        UT     & EXP&\multicolumn{2}{c}{\pv}&\multicolumn{2}{c}{\pq}&\multicolumn{2}{c}{\pu}&\multicolumn{2}{c}{\pl}&\multicolumn{2}{c}{$\Theta$}
                         &\multicolumn{2}{c}{\pv}&\multicolumn{2}{c}{\pq}&\multicolumn{2}{c}{\pu}&\multicolumn{2}{c}{\pl}&\multicolumn{2}{c}{$\Theta$}\\
&(s)&\multicolumn{2}{c}{(\%)}&\multicolumn{2}{c}{(\%)}&\multicolumn{2}{c}{(\%)}&\multicolumn{2}{c}{(\%)}&\multicolumn{2}{c}{($^\circ$)}&
                          \multicolumn{2}{c}{(\%)}&\multicolumn{2}{c}{(\%)}&\multicolumn{2}{c}{(\%)}&\multicolumn{2}{c}{(\%)}&\multicolumn{2}{c}{($^\circ$)}\\
\hline
2015-08-28 22:10 &1920& 3.39&0.01& \multicolumn{8}{c}{} & 3.99 & 0.00 & \multicolumn{8}{c}{} \\
2015-08-31 00:00 &2400& 3.37&0.02& \multicolumn{8}{c}{} & 3.99 & 0.01 & \multicolumn{8}{c}{} \\
2015-09-01 00:17 &2400&\multicolumn{2}{c}{}&1.67&0.08&2.70&0.08&3.17&0.08&29.1&0.7&\multicolumn{2}{c}{}&$-0.04$&0.08&$-0.33$&0.08&0.34&0.08&131.3&6.6\\ 
2018-09-22 21:51 &1800& 3.22&0.01& \multicolumn{8}{c}{} & 4.00 & 0.04 & \multicolumn{8}{c}{} \\
2018-09-22 21:01 &3600&\multicolumn{2}{c}{}&1.40&0.04&2.74&0.04&3.08&0.04&31.5&0.4&\multicolumn{2}{c}{}&$-0.10$&0.14&$-0.43$&0.14&0.44&0.14&128.6&9.0 \\
\hline
\end{tabular}
\end{small}
\end{table*}